%% file: main.tex
\documentclass[a4paper]{article}

\pdfoutput=1
\usepackage{lmodern}
\usepackage{booktabs}
\usepackage[T1]{fontenc}
\usepackage{subfloat}
\usepackage[latin9]{inputenc}
\usepackage{babel}

\makeatletter

\usepackage{mathrsfs}   %\mathscr{L}
\usepackage{slashed}     %\slashed{p}
\usepackage{url}
\usepackage{graphicx}
\usepackage{color} 
\usepackage{bm}
\usepackage[numbers,sort&compress]{natbib}
\usepackage{float}
\usepackage{multirow}
\usepackage{amsmath}
\usepackage{amssymb}
\usepackage{breqn}
\usepackage{subcaption}
\usepackage{caption}
\usepackage{mciteplus}
\usepackage{authblk}
\usepackage{diagbox}
\usepackage{stackengine}
\usepackage[skip=0pt]{caption} % change space between the figure and its caption 
\usepackage[normalem]{ulem}

\input{Packages/myTikz}
\input{Packages/myHypref}

\newcommand{\T}{\mathrm{T}}
\newcommand{\KaTie}{Ka\hspace{-0.2ex}Tie}

\begin{document}

\title{
 Heavy-hadron production based on $k_t$-factorization with scale-dependent fragmentation functions}
\author[]{F. E. Barattini}
\author[]{C. O. Dib}
\author[]{B. Guiot\thanks{benjamin.guiot@usm.cl} }
\affil[]{\it{\normalsize Departamento de F\'isica, Universidad T\'ecnica Federico Santa Mar\'ia; Casilla 110-V, Valparaiso, Chile}}

\renewcommand\Authands{ and }
\date{}

\maketitle
\begin{abstract}
We present a comprehensive study of heavy-hadron production, including $D$ and $B$ mesons, heavy baryons, and the $B_c$ meson. Our calculations are based on the $k_t$-factorization and scale-dependent fragmentation functions, completing the program of implementing this formalism within a variable-flavor-number scheme, as initiated in Ref.~\cite{Guiot2021}. Special emphasis is placed on the gluon-to-heavy-hadron contribution, which improves the description of data at small transverse momenta. Finally, we explore and compare different factorization schemes. We achieve the non-trivial task of providing a reasonable description of all the data considered in this study with a factorization scheme fixed once and for all.
\end{abstract}
\newpage

\tableofcontents

\section{Introduction \label{secintro}}

Factorization theorems allow a clear and relatively simple description of a certain number of observables involving hadrons. Their proof or formulation relies on approximations valid only in certain kinematical regions. In the small-$x$ regime with a sufficiently large hard scale, that is, $Q>Q_s$, with $Q_s$ the saturation scale, inclusive or semi-inclusive cross sections can be described by high-energy factorization (HEF) \cite{Catani1990,Catani1991,Collins1991,Levin91}, also known as $k_t$-factorization. Here, the cross section is given by the convolution of non-perturbative objects, the unintegrated parton distribution functions (UPDFs) and fragmentation functions (FFs), with an off-shell cross section, see also Eq.~(\ref{ktfac}). In the seminal papers, the UPDFs obey the Balitsky-Fadin-Kuraev-Lipatov (BFKL) equation \cite{Fadin1975,Kuraev1976,Kuraev1977,Balitsky1978}, resumming large logarithms of $x=Q^2/s$, where $\sqrt{s}$ is the center-of-mass energy of the colliding system. In principle, the HEF is not accurate for $Q<Q_s$ because it misses the genuine-twist corrections resumed to all order by the color-glass-condensate (CGC) effective theory. This formalism has been reviewed, for instance, in Refs.~\cite{Iancu2004,Gelis2010,Kovchegov2022}. The CGC can be applied to a larger kinematical region and contains the HEF, obtained from the former in the dilute-target approximation \cite{Kotko2015}. However, implementation of the HEF is simpler, and the saturation scale is relatively small for inclusive proton-proton collisions.
The relation between the CGC and the HEF and, more generally, between different small-$x$ formalisms is under active study. There are, in particular, numerous studies on the relation between the CGC and transverse-momentum-dependent (TMD) factorization, see \cite{Boussarie2023} and references therein. We could also mention the improved TMD factorization~\cite{Kotko2015}, bridging the HEF and TMD factorization.\\

%The HEF can be obtained in the dilute-target approximation (see, for instance, %Ref.~\cite{Kotko2015}) of the color-glass-condensate (CGC) effective theory, explaining %the presence of the saturation scale above. The CGC formalism has been reviewed, for %instance, in Refs.~\cite{Iancu2004,Gelis2010,Kovchegov2022}. In principle, the HEF is not %accurate for $Q<Q_s$ because it misses the genuine-twist corrections resumed to all order by the CGC. 

 To contribute to this effort from one side, and be able to provide precise predictions from the other side, it is essential to keep improving the HEF. Because this formalism has been developed for asymptotic energies, most calculations are based only on the unintegrated gluon density. However, present collider energies are not large enough to ignore the quark contribution~\cite{Guiot2019}. Despite contributing only beyond leading log accuracy in the HEF, the importance of the quark sector was already noted and discussed in detail in Ref.~\cite{Catani1994}. Moreover, including the contribution from initial heavy quarks is recommended for observables such as heavy-meson production at large transverse momentum. Consequently, one of our main goals is the development of a variable-flavor-number scheme (VFNS) for the HEF. In a VFNS with PDFs renormalized with the $\overline{\text{MS}}$ scheme, heavy-quark PDFs start contributing when the factorization scale is larger than the heavy-quark mass. In \cite{Guiot2021}, we performed HEF calculations using a VFNS for the UPDFs but used fixed FFs. The reason why it does not lead to a disastrous estimate of the heavy-flavor cross section is that the free parameter of, e.g., the Peterson FF \cite{Peterson1983}, can be chosen in such a way that it mimics the evolution. Of course, such a trick has limitations, and a better description of the data is expected by a full implementation of the FFs scale evolution.\\

The primary goal of the present work is to improve the calculations published in Ref.~\cite{Guiot2021} by including the scale evolution of FFs. We will pay particular attention to the  gluon contribution $D_{g\to H_Q}$, with $H_Q$ a heavy hadron. Then, we obtain for the first time results with the HEF in a full VFNS. In the framework of the HEF, the contributions $D_{g\to D}$ and $D_{g\to B}$ have been studied in Ref.~\cite{Karpishkov2015, Karpishkov2015a}. However, these calculations are not performed in a full VFNS because flavor-excitation contributions such as $gQ\to gQ$ are not considered. We will see that the main effect of the FF evolution is the improvement of theoretical results at small transverse momentum.\\

Our calculations, applied in \cite{Guiot2021} to $D$ and $B$ mesons, are extended to heavy baryons, demonstrating the potential of this formalism in describing a large number of observables. A global description of these data with a unified formalism is far from trivial. It is the reason why different schemes are sometimes employed depending on the specific observable. In this work, we try to identify the scheme that offers the best overall description of the data within our framework. Finally, we will present our predictions for the fragmentation contribution to the $B_c$ meson.

\section{Formalism}
\begin{figure}[t]
\begin{center}
 \includegraphics[width=22pc]{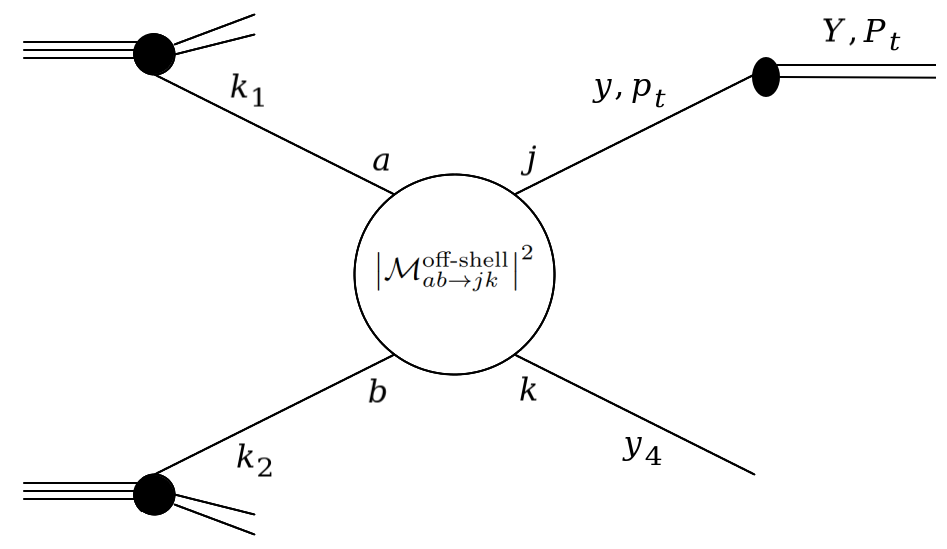}
\end{center}
\caption{Sketch of the leading-order pp collision with the hadronization of parton $j$ in the heavy hadron $H_Q$. $a$, $b$, $j$, and $k$ are the parton flavors, and the kinematics of parton $k$ is integrated out.\label{kine}}
\end{figure}

We work with the HEF, originally designed to resum large logarithms $\ln(s)$ in total cross sections \cite{Catani1990,Catani1991,Collins1991,Levin91}, with $\sqrt{s}$ the center of mass energy of the colliding system. We work at leading order, with the relevant variables presented in Fig.~\ref{kine}.
The proton-proton to parton differential cross section reads
\begin{multline}
 \frac{d\sigma(pp\to j + X)}{dyd^2p_t}(y,p_t;\mu_f)=\sum_{a,b}\int dy_4\int^{k_{\text{max}}^2}_0 dk^2_{1t}dk^2_{2t} F_{a}(x_1,k_{1t};\mu_i)\\
 \times F_{b}(x_2,k_{2t};\mu_i)d\hat{\sigma}_{ab\to j+X}(x_1,x_2,p_t,k_{1t},k_{2t};\mu_i,\mu_f), \label{ktfac}
\end{multline}
The variables $x_{1}$, $x_2$, $k_{1t}$, and $k_{2t}$ are the longitudinal momentum fractions and transverse momenta carried by the initial partons. The parton $j$ has rapidity $y$ and transverse momentum $p_t$, and there is an integration on $y_4$, the rapidity of the second outgoing parton. The variables $\mu_i$ and $\mu_f$ are the factorization and fragmentation scales, respectively. The functions $F_a$ are the unintegrated PDFs (UPDFs), and $d\hat{\sigma}_{ab\to j+X}$ is the (differential) off-shell partonic cross section
\begin{equation}
    d\hat{\sigma}_{ab\to j+X}=\frac{1}{16\pi^2\hat{s}^2}\sum_k \left|\mathcal{M}^{\text{off-shell}}_{ab\to jk} \right|^2\frac{1}{1+\delta_{jk}},
\end{equation}
where the sum runs on all possible parton flavor $k$, and $\hat{s}=x_1x_2s$. In the equation above, the usual delta function encoding 4-momentum conservation is absent because it is used in Eq.~(\ref{ktfac}) to integrate out some of the momenta. The $\delta_{jk}$ takes care of final identical particles, and the initial partons $a$ and $b$ are off-shell, with $k_1^2=-k^2_{1t}$ and $k_2^2=-k^2_{2t}$. The treatment of off-shell matrix elements requires special techniques which have been exposed, for instance, in \cite{Hameren2013}, and the cross section of Eq.~(\ref{ktfac}) is obtained with the help of the event generator \KaTie\  \cite{Hameren2018}. To facilitate the comparison with the present work, we use the UPDFs built in Ref.~\cite{Guiot2021}. These are based on the Kimber-Martin-Ryskin-Watt (KMRW) prescription \cite{Kimber2001,Watt2003} and obey approximately \footnote{See Ref.~\cite{Guiot2023} for more details on this point.}
\begin{equation}
    xf_a(x;\mu)=\int_0^{\mu^2}F_a(x,k_t^2;\mu)dk_t^2, \label{norm}
\end{equation}
where $f_a$ are the collinear PDFs. At all orders, the l.h.s of Eq.~(\ref{ktfac}) is independent of the scale $\mu_i$. On the opposite, it depends on the fragmentation scale $\mu_f$. This dependence disappears in the heavy-hadron cross section, obtained by a convolution of Eq.~(\ref{ktfac}) with FFs
\begin{multline}
    \frac{d\sigma(pp\to H_Q + X)}{dYd^2P_t}(Y,P_t)=\sum_j\int_{z_{\text{min}}}^1 \frac{dz}{z^2}D_{j\to H_Q}(z;\mu_f)\\
    \times \frac{d\sigma(pp\to j + X)}{dyd^2p_t}\left(y,\frac{P_t}{z};\mu_f\right). \label{facfrag}
\end{multline}
Here, $Y$ and $P_t$ are the rapidity and transverse momentum associated with the heavy hadron $H_Q$, where the index $Q$ indicates that it is made of a heavy quark of flavor $Q=c,b$. The relation between the parton and hadron rapidities is entirely determined by the definition of the fragmentation variable, which, in our case, is
\begin{equation}
    \vec{P}=z\vec{p}.
\end{equation}
This leads to the expression
\begin{equation}
    y=\ln \left(\frac{E_{\text{parton}}+P_z/z}{\sqrt{m^2+P_t^2/z^2}} \right), \label{relyy}
\end{equation}
with $E_{\text{parton}}$ obtained from the on-shell condition, and
\begin{equation}
    P_z=M_t\sinh(Y),
\end{equation}
with $M_t$ the heavy-hadron transverse mass. For $D$-meson production, the effect on the cross section due to the shift in rapidity is negligible, even for gluons with $m=0$.\\

In practice, we work at leading order, and the heavy-hadron cross section retains a dependence on $\mu_i$ and $\mu_f$ leading to the usual scale uncertainties. Our default choice is
\begin{equation}
    \mu_i=m_t=\sqrt{m^2+p_t^2}, \quad \mu_f=M_t, \label{factscale}
\end{equation} 
however, we will also explore other possibilities. This choice of factorization and fragmentation scales, while perfectly valid, differs from a standard parametrization
\begin{equation}
    \mu_i=\mu_f=M_t,
\end{equation}
found in the literature. We note that Eq.~(\ref{factscale}) makes sense from a spacetime point of view. Indeed, the initial radiations, associated with the scale $\mu_i$ and emitted long before the hadronization process, are blind to $P_t=zp_t$.

\section{$D$-meson production}
\subsection{Evolution and prediction for central rapidity}
Our earlier work \cite{Guiot2021} focused exclusively on cases where the parton $j$ of Eq.~(\ref{facfrag}) is a charm or bottom quark. However, it receives contributions from the fragmentation of light flavors to heavy hadrons when $\mu_f>\mu_0$, with $\mu_0$ the initial scale for the (timelike) Dokshitzer-Gribov-Lipatov-Altarelli-Parisi (DGLAP) evolution \cite{GribovLipatov:1972,AltarelliParisi:1977,Dokshitzer:1977}. In this study, we include the gluonic term $D_{g\to H_Q}$. The role of other light flavors is generally negligible.
The DGLAP evolution of our fragmentation functions is performed with the help of QCDNUM \cite{Botje2011}, and we fix the initial scale at $\mu_0=m_Q$, with $m_Q$ the heavy-quark mass. In this section, we work with the factorization and fragmentation scales defined in Eq.~(\ref{factscale}).\\

\begin{figure}[h!]
\begin{center}
 \includegraphics[width=22pc]{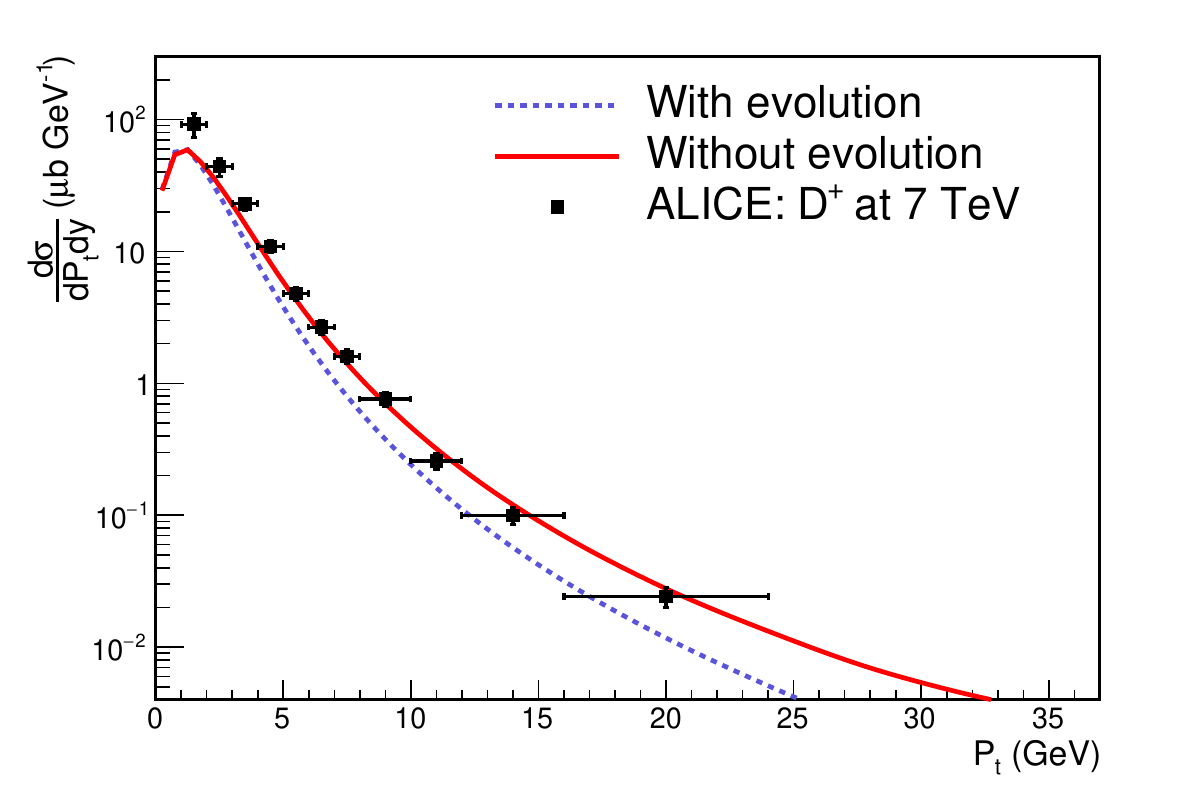}
\end{center}
\caption{Comparison of our calculations, Eq.~(\ref{facfrag}), with ALICE data \cite{Acharya2017} at 7 TeV and $|y|<0.5$. The blue and red lines correspond to the evolved and fixed Peterson FF, respectively.\label{Al7}}
\end{figure}

We start our analysis with $D^+$ data at 7 TeV. In Fig.~\ref{Al7}, we compare our calculations using evolved (dashed blue line) and fixed (red line) FFs with ALICE data  \cite{Acharya2017}.
\begin{figure}[h!]
\begin{center}
 \includegraphics[width=22pc]{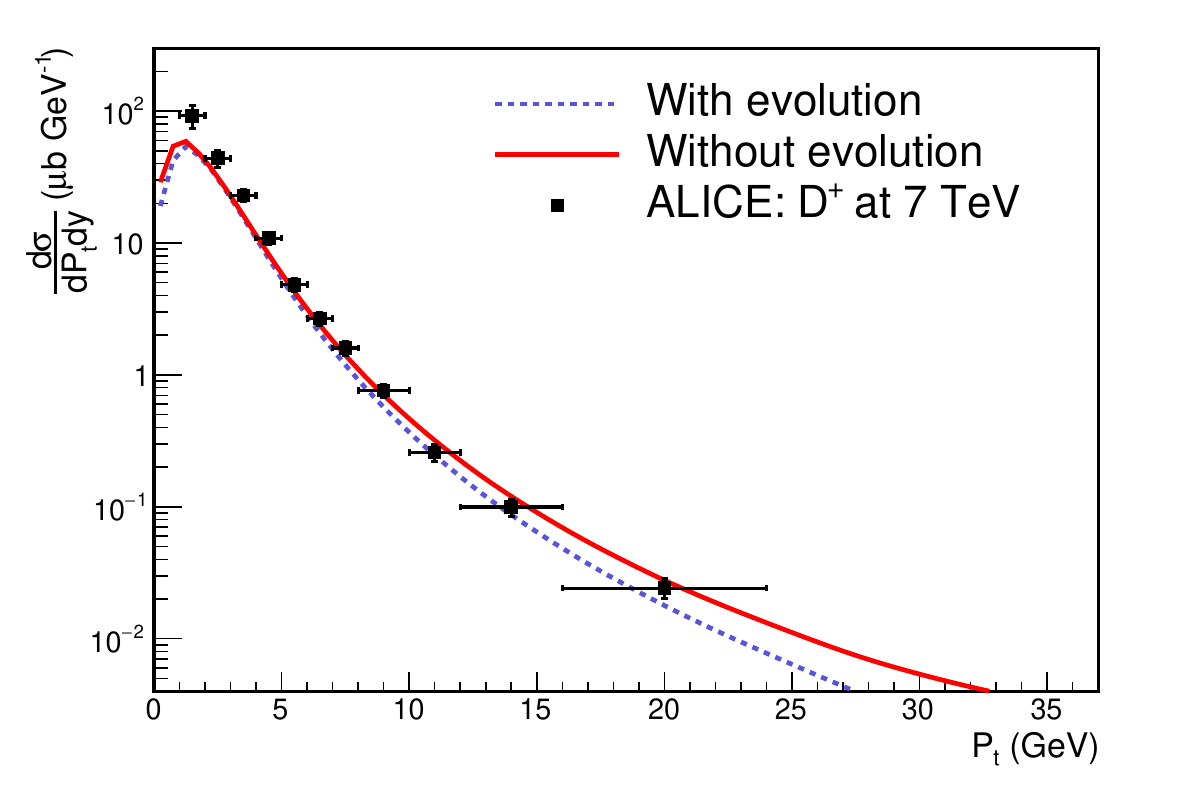}
\end{center}
\caption{Same as Fig.~\ref{Al7}, except that we switched $\epsilon_c$ from 0.05 to 0.008 for the evolved FF (blue line). \label{Dp7new}}
\end{figure} 
We used the Peterson FF
\begin{equation}
    D_{i\to H_Q}(z)= f(i\to H_Q)N(\epsilon_i)\frac{1}{z\left(1-\frac{1}{z}-\frac{\epsilon_i}{1-z}\right)^2},
\end{equation}
with $N(\epsilon_i)$ the normalization factor. For the $c\to D^+$ case, we set $\epsilon_c=0.05$, and use the standard fragmentation fraction $f(c\to D^+)=0.234$. As expected, the effect of evolution is visible at large $P_t$, while negligible at small transverse momentum. Apparently, the evolution worsens the result. The reason is that the parameter $\epsilon_c$ is extracted from data by a fit that does not include the fragmentation-scale evolution. However, it is possible to adjust $\epsilon_c$ to obtain a satisfactory description of the data by the evolved FF. In Fig.~\ref{Dp7new}, we show the result obtained with $\epsilon_c=0.008$ for the evolved FF, while keeping $\epsilon_c=0.05$ for the fixed FF. The difference between the two FFs is reduced, and we conclude that an appropriate choice of the free parameter mimics the evolution of the $D_{c\to D}$ function.\\

\begin{figure}[t]
\begin{center}
 \includegraphics[width=22pc]{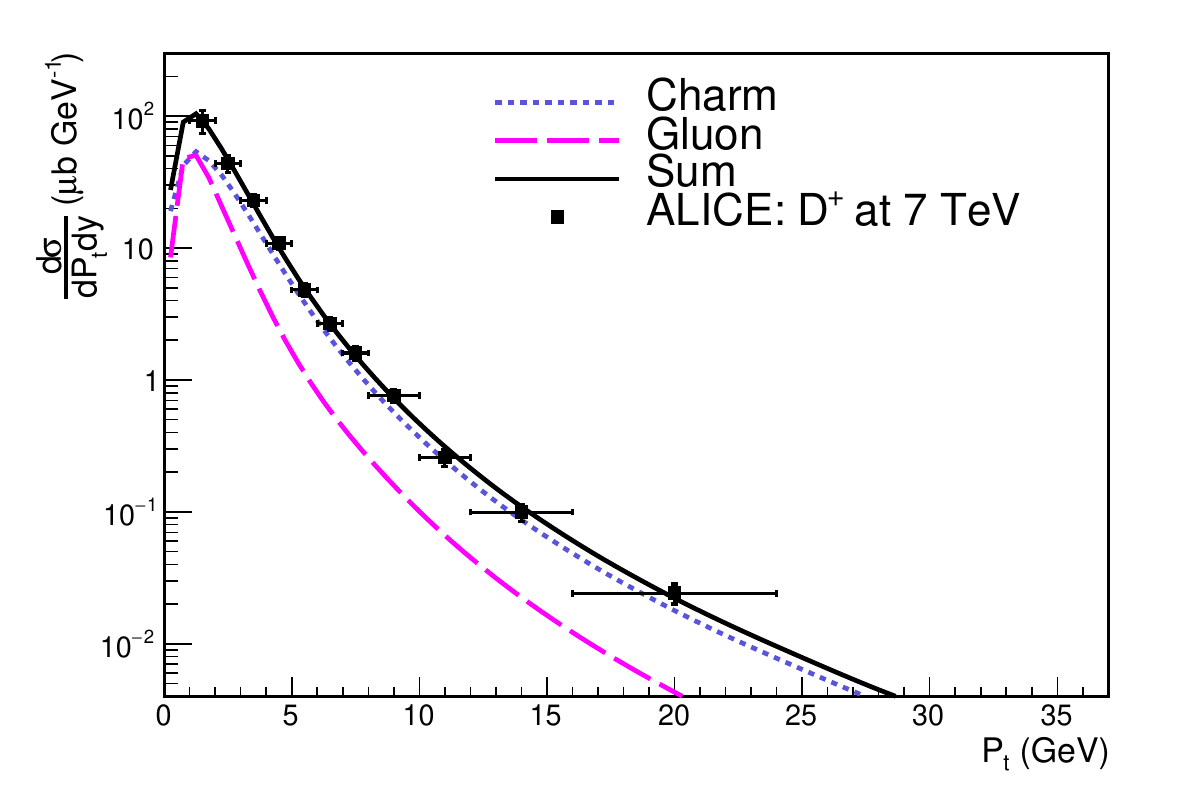}
\end{center}
\caption{Differential cross section including charm and gluon fragmentation to $D^+$ meson. The total contribution (black line) is compared to ALICE data \cite{Acharya2017} at 7 TeV and $|y|<0.5$.\label{Dpf}}
\end{figure}

A key feature of scale-dependent FFs is the contribution of partons other than a charm quark, particularly the gluon contribution $D_{g\to D^+}$. Including both $g\to D^+$ and $c\to D^+$ channels, we obtain the result shown in Fig.~\ref{Dpf}.
\begin{figure}[t!]
\begin{center}
 \includegraphics[width=22pc]{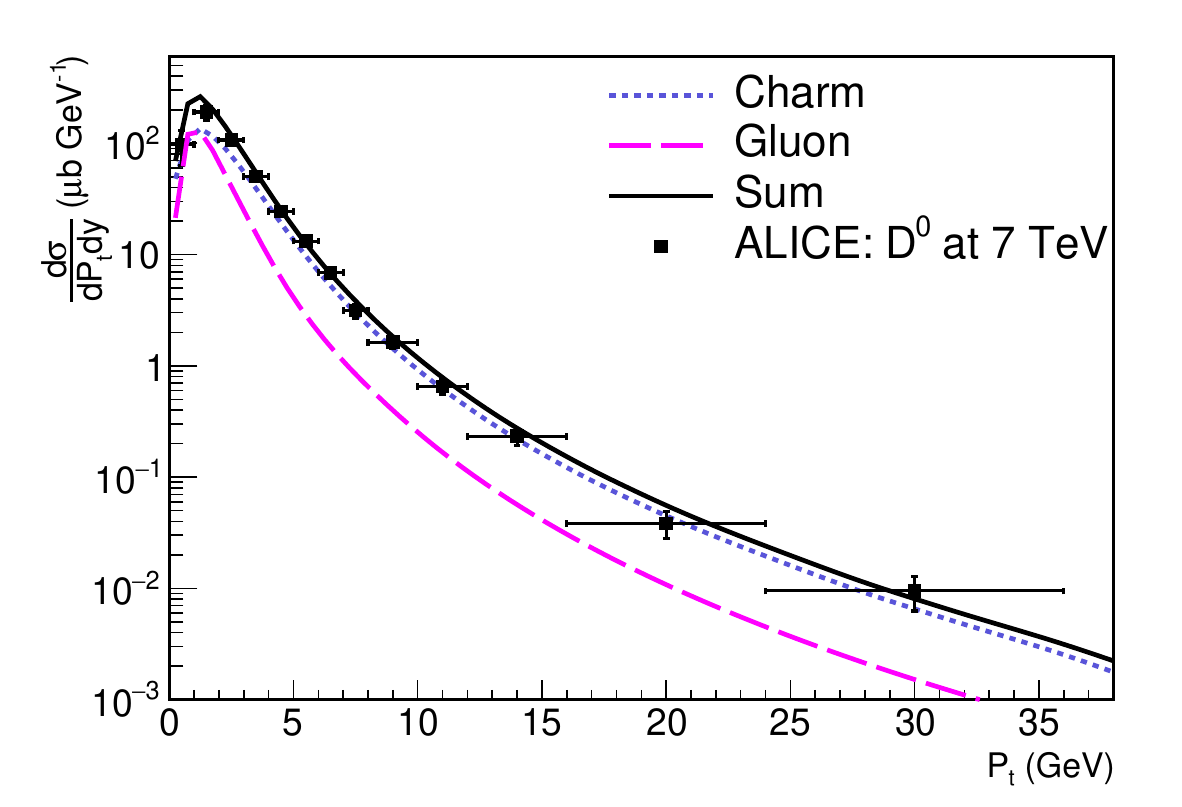}
\end{center}
\caption{Differential cross section including charm and gluon fragmentation to $D^0$ meson. The total contribution (black line) is compared to ALICE data \cite{Acharya2017} at 7 TeV and $|y|<0.5$.\label{D07tev}}
\end{figure}
At small transverse momentum, there is a clear improvement of the result obtained by the charm contribution alone. Above $P_t \sim 2 m_c$, the gluon contribution is negligible. We show a similar result for the $D^0$ meson in Fig.~\ref{D07tev}.\\

\begin{figure}[h!]
\begin{center}
 \includegraphics[width=22pc]{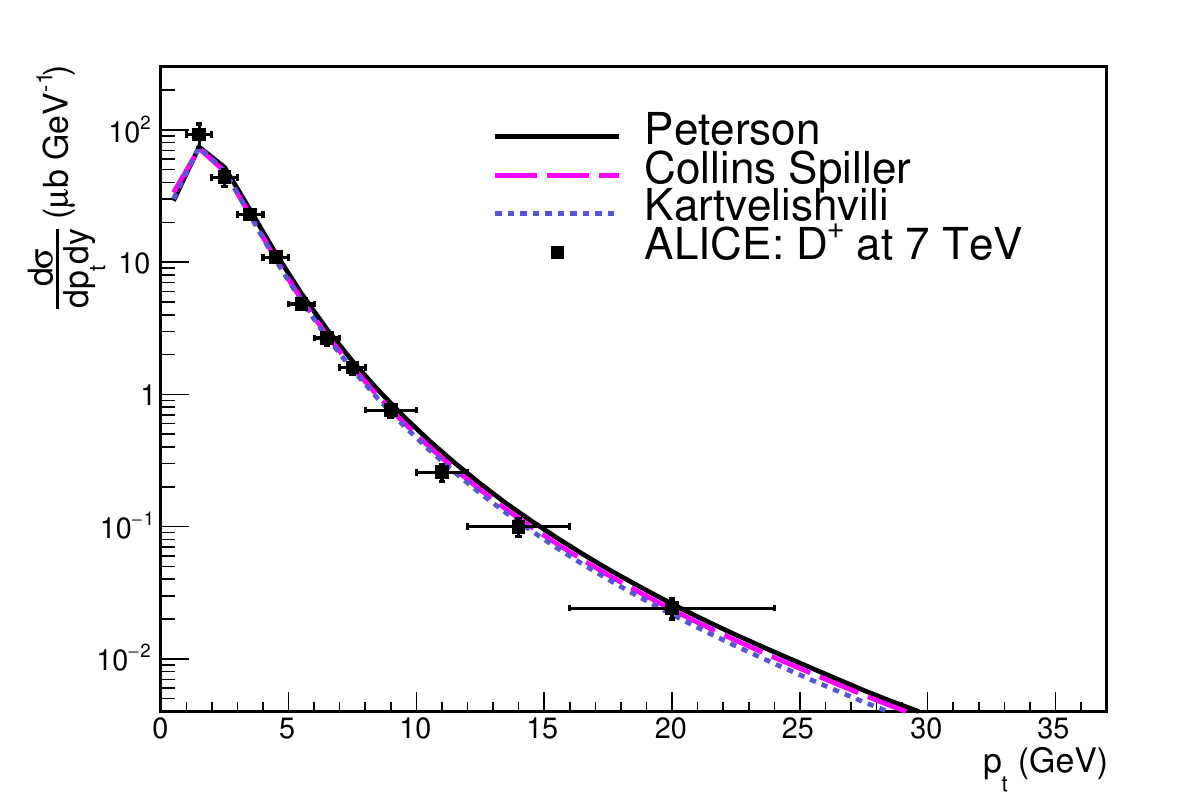}
\end{center}
\caption{Comparison of different FFs, see the text for more details. The differential cross section is compared to ALICE measurement \cite{Acharya2017}. \label{frag}}
\end{figure}

We close this section by comparing in Fig.~\ref{frag} the results obtained with different fragmentation functions. Namely, we used the Peterson and Collins-Spiller FFs \cite{Collins1985} with $\epsilon_c=0.008$ and the Kartvelishvili FF model \cite{Kartvelishvili1978} with $\alpha_c=7$. Remember that these values may differ from the ones encountered in the literature because of the FF evolution. With the appropriate choice of the non-perturbative parameter, all these FFs give similar results.

\subsection{The BKKSS and SACOT-$m_T$ schemes}
\begin{figure}[h!]
\begin{center}
 \includegraphics[width=22pc]{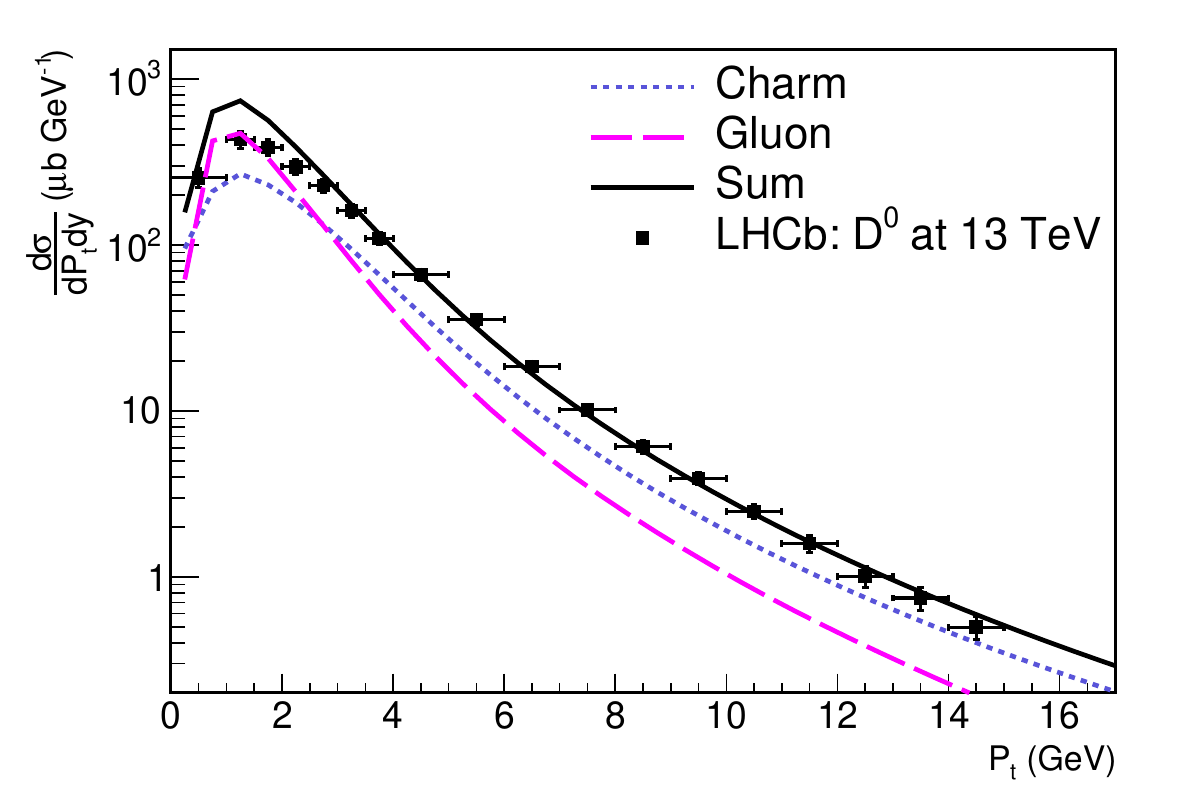}
\end{center}
\caption{Comparison of our calculations using the default scales, Eq.~(\ref{factscale}), with LHCb data \cite{Aaij2016} for the rapidity range $2<y<2.5$. \label{D0225}}
\end{figure}

Turning to $D$-meson production at forward rapidity, we encounter the issue visible in Fig.~\ref{D0225}. The gluon contribution at small $P_t$ is too large, and our result overestimates the data.
The total contribution is still in reasonable agreement with the data, accounting for the large uncertainties in the small-$P_t$ region. However, the situation gets worse for $B$ mesons, and we prefer to look immediately for a solution. Ideally, we aim to find a scheme working for all heavy hadrons discussed in this paper. In this section, we will try the BKKSS and SACOT schemes.\\

By BKKSS, we refer to the factorization scale used by Benzke, Kniehl, Kramer, Schienbein, and Spiesberger in \cite{Benzke2019} for $B$ mesons, i.e.,
\begin{equation}
    \mu_f=0.49\sqrt{P_t^2+4 m_Q^2}. \label{mup}
\end{equation}
Following the logic of Eq.~\eqref{factscale}, we set the factorization scale to
\begin{equation}
    \mu_i=0.49\sqrt{p_t^2+4 m_Q^2}, \label{mup2}
\end{equation}
while Ref.~\cite{Benzke2019} used $\mu_i=\mu_f$.
Our second choice has been inspired by the SACOT-$m_\T$ scheme (SACOT stands for Simplified Aivazis-Collins-Olness-Tung) \cite{Helenius2018,Helenius2023}. Here, the so-called non-direct channels, including the $gg\to gg$ process, are computed in a zero-mass scheme. However, in the factorization formula (\ref{ktfac}), the partonic cross sections associated with these processes are evaluated at $m_t$ rather than $p_t$. In that case, we keep using the scales defined in Eq.~(\ref{factscale}). The first interest of this scheme is that it regularizes the divergences of massless-parton cross sections when $p_t \to 0$. It is also consistent with our choice for the factorization and fragmentation scales, given by Eq.~(\ref{factscale}), involving $m_Q$. This implementation can be viewed as accounting for the mass effect in the splitting $g\to Q\bar{Q}$ necessary for the fragmentation of a light parton into a heavy meson.
\begin{figure}[t!]
\begin{center}
 \includegraphics[width=22pc]{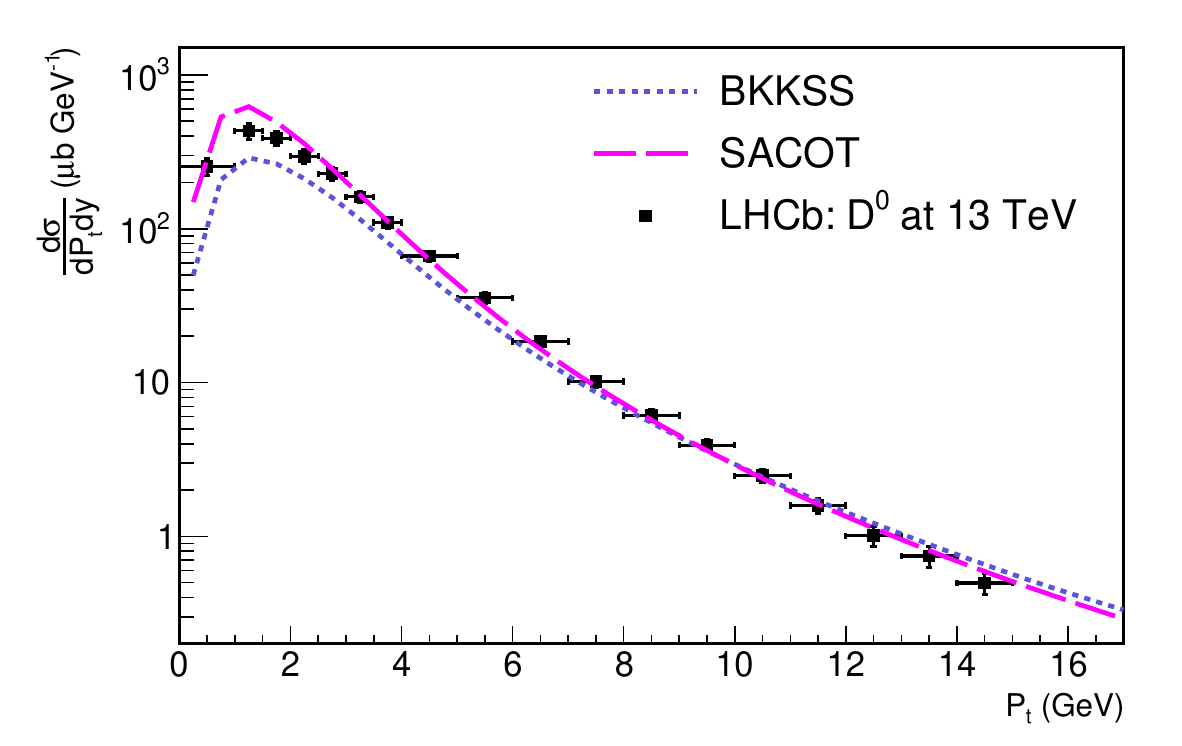}
\end{center}
\caption{Comparison of our calculation using the BKKSS and SACOT-$m_\T$ schemes, with LHCb data \cite{Aaij2016} for the rapidity range $2<y<2.5$. \label{comp225_f}}
\end{figure}
we show the results obtained at forward rapidity in Fig.~\ref{comp225_f}. While the SACOT-$m_\T$ central prediction improves the result of Fig.~\ref{D0225}, the BKKSS scheme slightly underestimates the cross section at small $P_t$. We checked that the SACOT-$m_\T$ scheme also leads to a satisfying result for central rapidities. It was expected because the prescription for non-direct channels affects mainly the gluon-to-heavy-hadron contribution, which dominates at small $P_t$ and forward rapidity, but is similar or below the charm contribution at central rapidity. Consequently, the correction at small $P_t$ is just a few percent.

\section{$B$-meson production \label{secbmes}}
\begin{figure}[h]
\begin{center}
 \includegraphics[width=22pc]{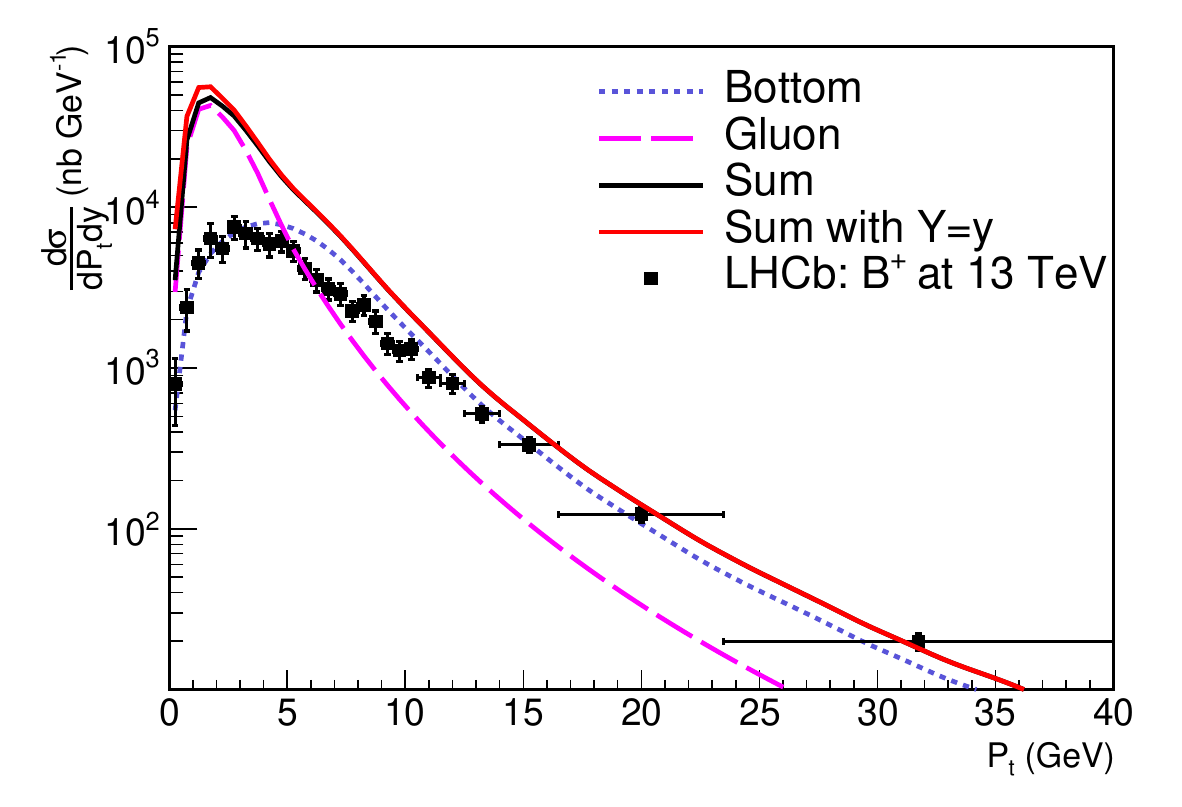}
\end{center}
\caption{Theoretical results obtained with the initial scheme compared to $B$-meson data measured at 13 TeV and $2<y<2.5$ by the LHCb collaboration \cite{Aaij2017b}.\label{bmes}}
\end{figure}

The case of $B$-meson production is more complicated because of its large mass and due to the gluon contribution. In the following, we use the Peterson fragmentation function with the fragmentation fraction $f(b\to B)=0.403$ \cite{Cacciari2012} and the parameter $\epsilon_B=0.001$. Here, $B$ refers to $B^+$ or $B^0$ mesons (and associated antiparticles). Using our default scheme, i.e., without the SACOT procedure and with the scales set by Eq.~(\ref{factscale}) , we observe from Fig.~\ref{bmes} a significant overestimation of the cross section by the gluon fragmentation contribution at small $P_t$. Additionally, we show in Fig.~\ref{bmes} (red line) the total contribution obtained when we set the parton and hadron rapidities equal. We see that the more rigorous treatment given by Eq.~(\ref{relyy}) amounts only to a small decrease in the theoretical result. This behavior can be explained by the decrease in the cross section as a function of $y$ and $y>Y$, with $y$ the parton rapidity.\\

\begin{figure}[h!]
\begin{center}
 \includegraphics[width=22pc]{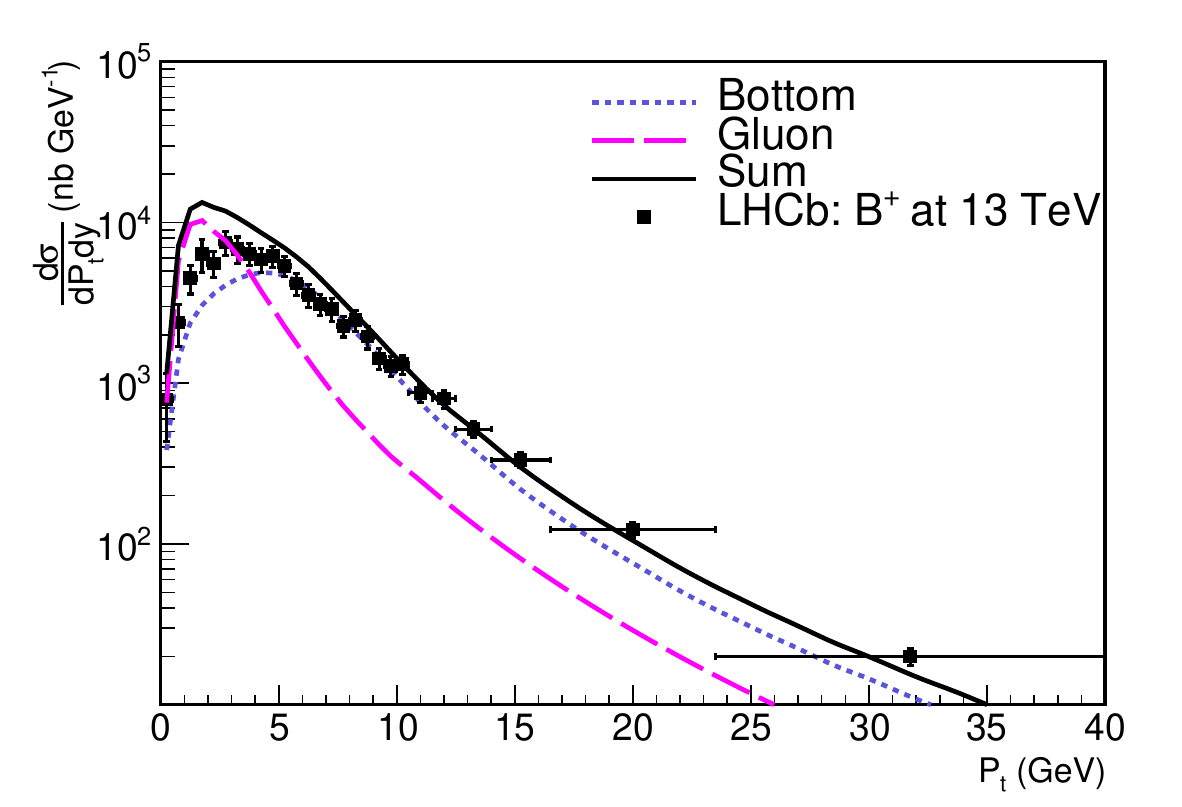}
\end{center}
\caption{Theoretical results obtained with the BKKSS scheme compared to $B$-meson data measured at 13 TeV and $2<y<2.5$ by the LHCb collaboration \cite{Aaij2017b}. \label{bmes2}}
\end{figure}

While $m_t\sim p_t$ in charm production, it is not the case anymore for the bottom quark with $m_b=4.75$ GeV, at least for $p_t< 10$ GeV. Consequently, the result is more sensitive to the scheme used in the calculation. There is, for instance, a significant improvement with the BKKSS scheme defined in Eqs.~(\ref{mup})-(\ref{mup2}), see Fig.~\ref{bmes2}.
We observe that the bottom contribution is now slightly below the data. This is an initial-state effect due to a smaller value of the factorization scale in the UPDFs. This contribution is poorly affected by the change of the fragmentation scale $\mu_f$. It is exactly the opposite situation for the gluon contribution, with a strong dependence on $\mu_f$. It was to be expected since $F_b(x,k_t;\mu_i=m_b)=0$ and $D_{g\to B}(z;\mu_f=m_b)=0$. These two distributions increase quickly when the scale grows larger than the initial scale $\mu_0=m_b$, due to their $\ln(\mu/m_b)$ dependence. The data are still overestimated at small $P_t$, but this result is certainly fine within scale uncertainties.\\

\begin{figure}[t!]
\begin{center}
 \includegraphics[width=22pc]{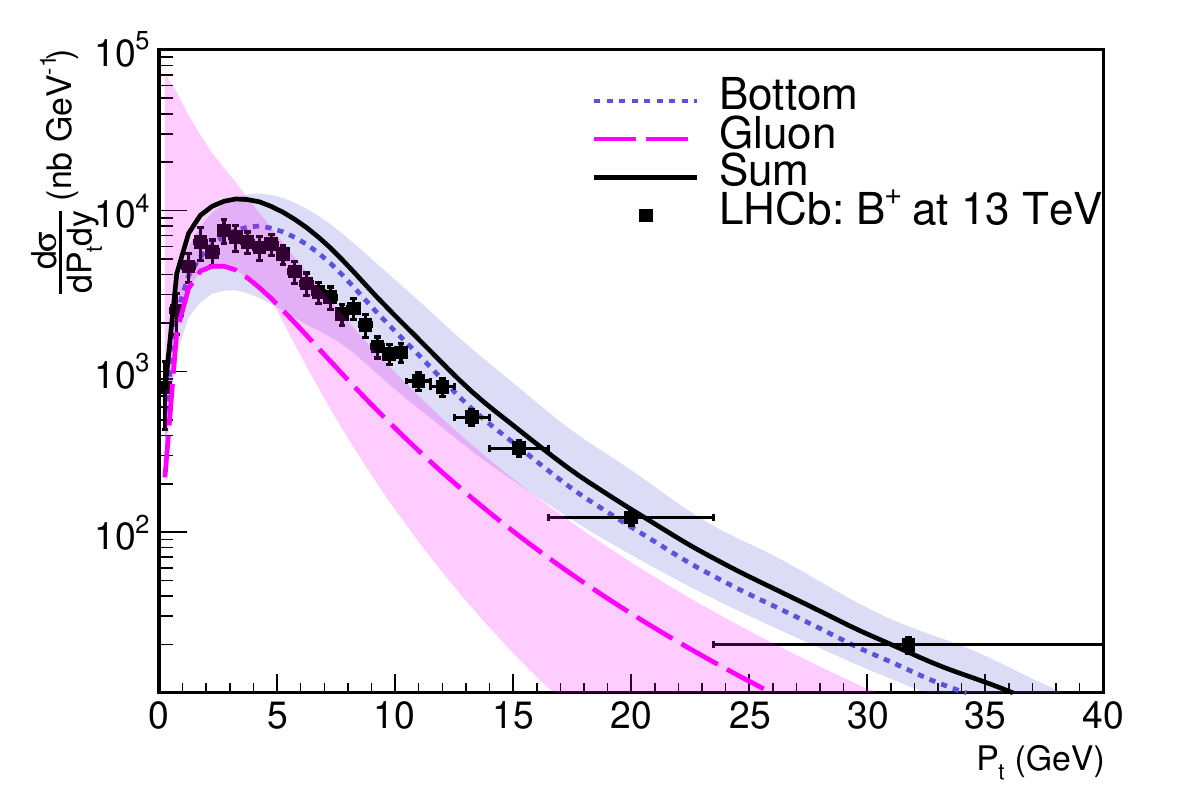}
\end{center}
\caption{Comparison of our calculations using the SACOT-$m_\T$ prescription for massless partonic cross sections and scales defined in Eq.~(\ref{factscale}) with LHCb data at 13 TeV and $2<y<2.5$ \cite{Aaij2017b}. \label{bmt}}
\end{figure}

Another interesting choice is the SACOT-$m_\T$ scheme presented in the previous section. A comparison between our calculations and LHCb data is shown in Fig.~\ref{bmt}.
The shape looks better, but, compared to the BKKSS scheme, the decrease of the gluon contribution with $P_t$ is slower, leading to a small overestimation at intermediate transverse momentum. In Fig.~\ref{bmt}, we include a rough (and conservative\footnote{There is a freedom in the estimation of scale uncertainties. Because NLO scale uncertainties are smaller compared LO, it is interesting to use a conservative estimation in the latter case.}) estimate of the scale uncertainties by varying $\mu_i$ by a factor of $\sqrt{2}$ for the bottom contribution and $\mu_f$ by the same amount for the gluon contribution. $\mu_f$ and $\mu_i$ are kept fixed for the bottom and gluon contributions, respectively. As discussed earlier, the dependencies of these two contributions on these scales are weaker. There is no lower uncertainty for the gluon contribution until $\mu_f=m_t/\sqrt{2}>\mu_0$, i.e., when $p_t=m_b$. With increasing transverse momentum, the phase space opens and the uncertainty band grows from below. To keep the figure readable, we do not show the total uncertainty band.

\section{Results with the new KMRW UPDFs}
\begin{figure}[t]
\begin{center}
 \includegraphics[width=22pc]{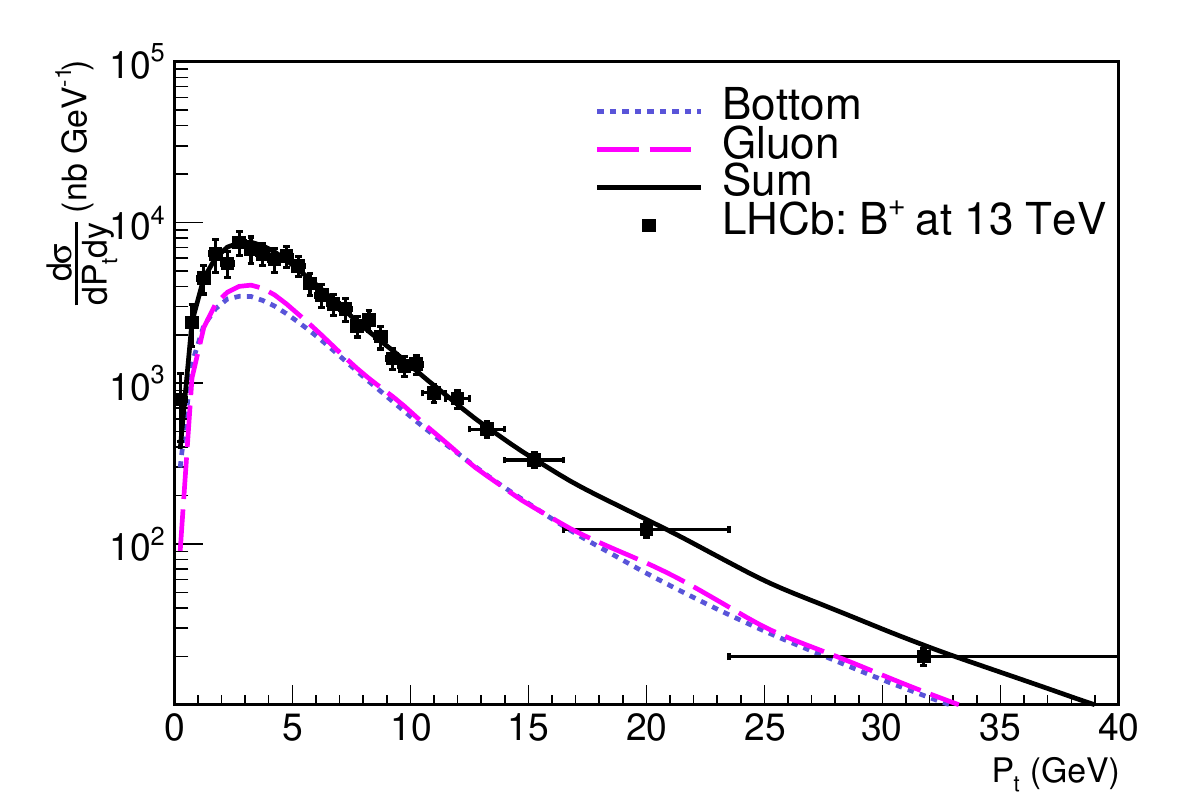}
\end{center}
\caption{Same as Fig.~\ref{bmt} using the modified KMRW UPDFs.\label{newmt2}}
\end{figure}

In this section, we briefly study the dependence of our earlier results on the UPDFs set. We use the modified KMRW (MK) UPDFs built in Ref.~\cite{Guiot2023}. While the usual KMRW UPDFs obey approximately Eq.~(\ref{norm}), the MK set has the exact normalization
\begin{equation}
    xf_a(x;\mu)=\int_0^\infty F_a(x,k_t^2;\mu)dk_t^2.
\end{equation}
The implementation of the MK UPDFs is simple and solves several issues of the original UPDFs, listed in \cite{Guiot2021}. This set should be used with $k^2_{\text{max}}$, defined in Eq.~(\ref{ktfac}), larger than the hard scale,\footnote{A detailed discussion on $k^2_{\text{max}}$ and its relation to the definition of UPDFs can be found in Ref.~\cite{Guiot2024}.} which is the standard choice in HEF calculations. The result for $B$-meson production with the scales from Eq.~(\ref{factscale}) and the SACOT-$m_\T$ scheme is presented in Fig.~\ref{newmt2}, and can be compared with Fig.~\ref{bmt}.
The surprising similarity between the gluon and bottom contributions is a coincidence. The sum of these two contributions is in good agreement with the data, while the central prediction of Fig.~\ref{bmt} was above by a factor of 2 or less at small $P_t$.\\

\begin{figure}[h!]
\begin{center}
 \includegraphics[width=22pc]{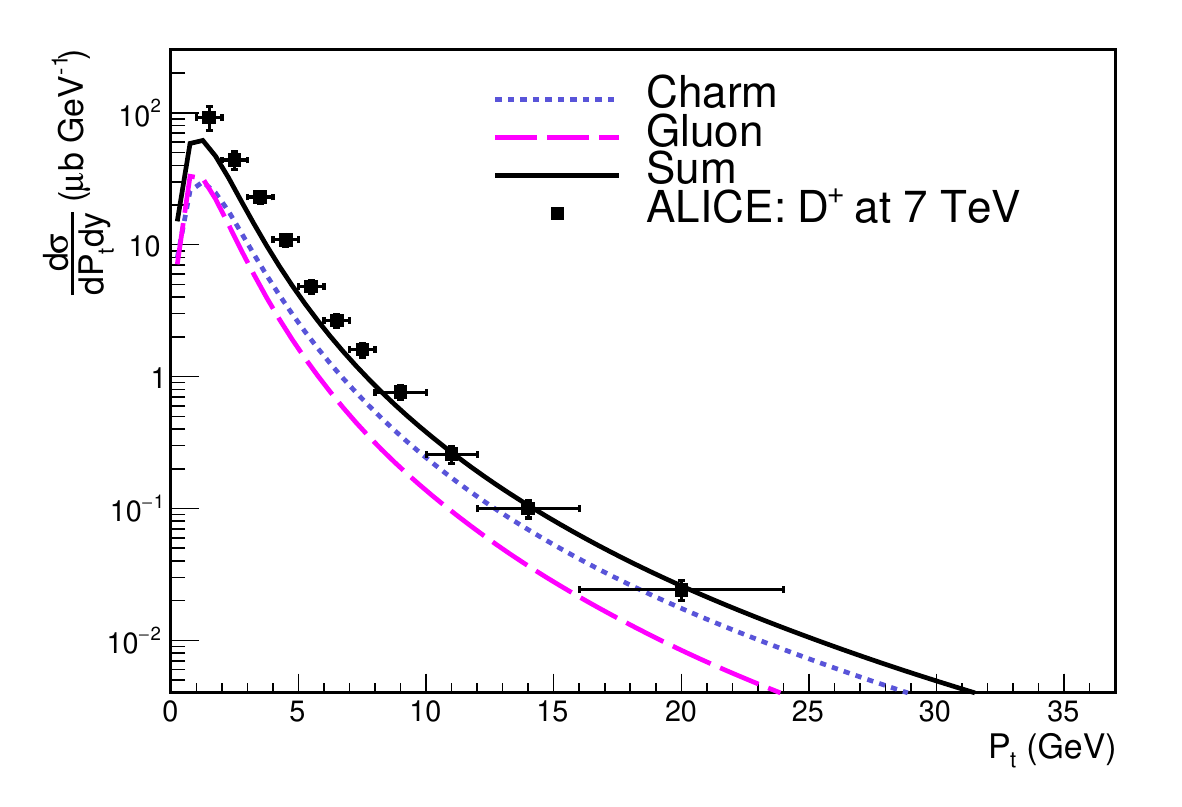}
\end{center}
\caption{$D^+$ transverse-momentum distribution at $|y|<0.5$ and $\sqrt{s}=7$ TeV measured by the ALICE collaboration \cite{Acharya2017} compared to our calculations for charm and gluon fragmentation. We used the SACOT-$m_\T$ scheme and the scales from Eq.~(\ref{factscale}).\label{cmtnew}}
\end{figure}
Using the same setup, we show in Fig.~\ref{cmtnew} the result for the $D^+$ meson. Compared to Fig.~\ref{Dpf}, and leaving aside considerations on scale uncertainties, our central result underestimates the data for $P_t<10$ GeV. The conclusion that the gluon fragmentation helps to describe $D$-meson data at very small $P_t$ still holds. We also conclude that the SACOT-$m_\T$ scheme allows a global description of $D$ and $B$ mesons at central and forward rapidities, while our more standard calculations failed due to a too large gluon contribution at forward rapidity. Finally, we observe that a significant theoretical uncertainty arises from the choice of UPDFs.

%But the underestimation is smaller compared to the case without evolution, see Ref.~\cite{Guiot2021} or the blue curve of Fig.~\ref{Dpf}. Then, the conclusion that the gluon fragmentation helps to describe $D$-meson data at very small $p_t$ still holds. We also conclude that the SACOT-$m_\T$ scheme allows a global description of $D$ and $B$ mesons at central and forward rapidities, while our more standard calculations failed due to a too large gluon contribution at forward rapidity.

\section{Heavy baryons}
\begin{figure}[h]
\begin{center}
 \includegraphics[width=24pc]{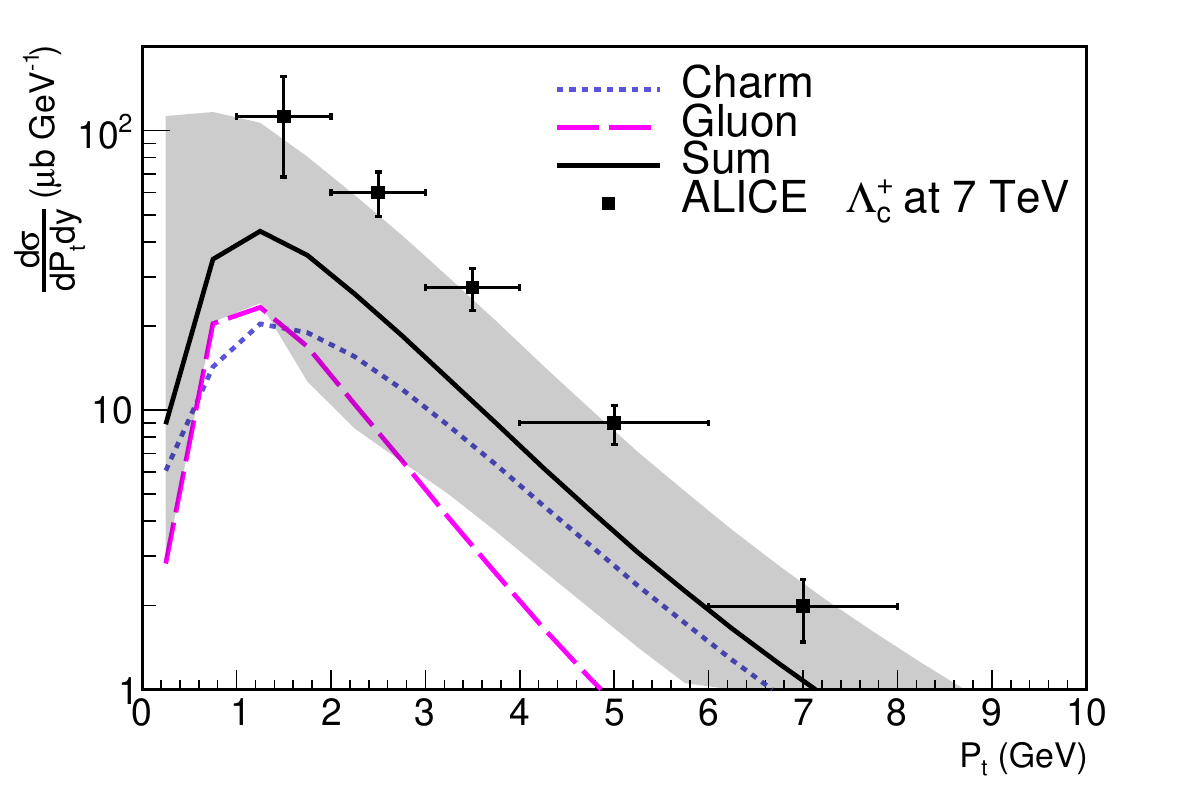}
\end{center}
\caption{Comparison between our calculations at 7 TeV in the rapidity range $|y|<0.5$ with ALICE data \cite{Acharya2018}.\label{lcAlice}}
\end{figure}

The production of heavy baryons presents a challenge for perturbative QCD. It appears that the hadronization process cannot be described solely by FFs at small transverse momentum. In the 2020 paper \cite{Kniehl2020}, the authors showed that their NLO calculations provided a reasonable description of LHCb ($2\leq y \leq 4.5$; $2 \leq P_t\leq 8$ GeV) and CMS ($|y|\leq 1$; $5\leq P_t\leq 20$ GeV) data \cite{Aaij2013,Sirunyan2020}, while underestimating ALICE data \cite{Acharya2018} ($|y|\leq 0.5$; $1\leq P_t \leq 8$ GeV). Our result for the production of $\Lambda_c^+$ baryons is compared to ALICE measurement at 7 TeV \cite{Acharya2018} in Fig.~\ref{lcAlice}.
We used the same setup as for Fig.~\ref{Dpf}, in particular, with the scales fixed by Eq.~(\ref{factscale}). In this section, we will not use the SACOT scheme, which, for central rapidities, does not affect significantly the total contribution. The (conservative) scale uncertainties associated with the gluon and charm contributions, described in Sec.~\ref{secbmes}, were added in quadrature to obtain the black error band. Following Ref.~\cite{Maciula2018}, we set the fragmentation fraction to 0.1.\footnote{This value is a bit larger than the one used in Ref.~\cite{Kniehl2020}, but our calculations are LO. In general, higher-order hard coefficients imply smaller non-perturbative functions (as long as the higher-order corrections are positive). In principle, this value already includes the contribution from charm exited-state decays, see \cite{Cacciari2003} for a detailed discussion on $D$ mesons.} We use again the Peterson FF, with the non-perturbative parameter $\epsilon_{\Lambda_c}=0.001$, i.e., the value used for $B$ mesons. In agreement with Refs.~\cite{Maciula2018,Kniehl2020}, our prediction underestimates the data.\\

At the time of Ref.~\cite{Kniehl2020}, it was unclear whether ALICE and CMS experiments were compatible, the latter being reasonably well described by calculations. One possible explanation for the discrepancy was the larger rapidity bin in the CMS experiment. Another possibility was an enhancement of the production at small $p_t$ and central rapidity. Since then, CMS published new results with an extended range $P_t\in[3,30]$ GeV \cite{Tumasyan2024}. At low transverse momentum, the two experiments seem to show compatible cross sections, bearing in mind that the CMS measurement was performed at 5 TeV. As expected, the comparison with GM-VFNS calculations demonstrates a significant underestimation of CMS data at low transverse momentum. On the opposite, PYTHIA8 calculations equipped with the mechanism of color reconnection \cite{Christiansen2015}, increasing the production of $\Lambda_c$ at low $p_t$, describe the data better.\\

\begin{figure}[h!]
\begin{center}
 \includegraphics[width=24pc]{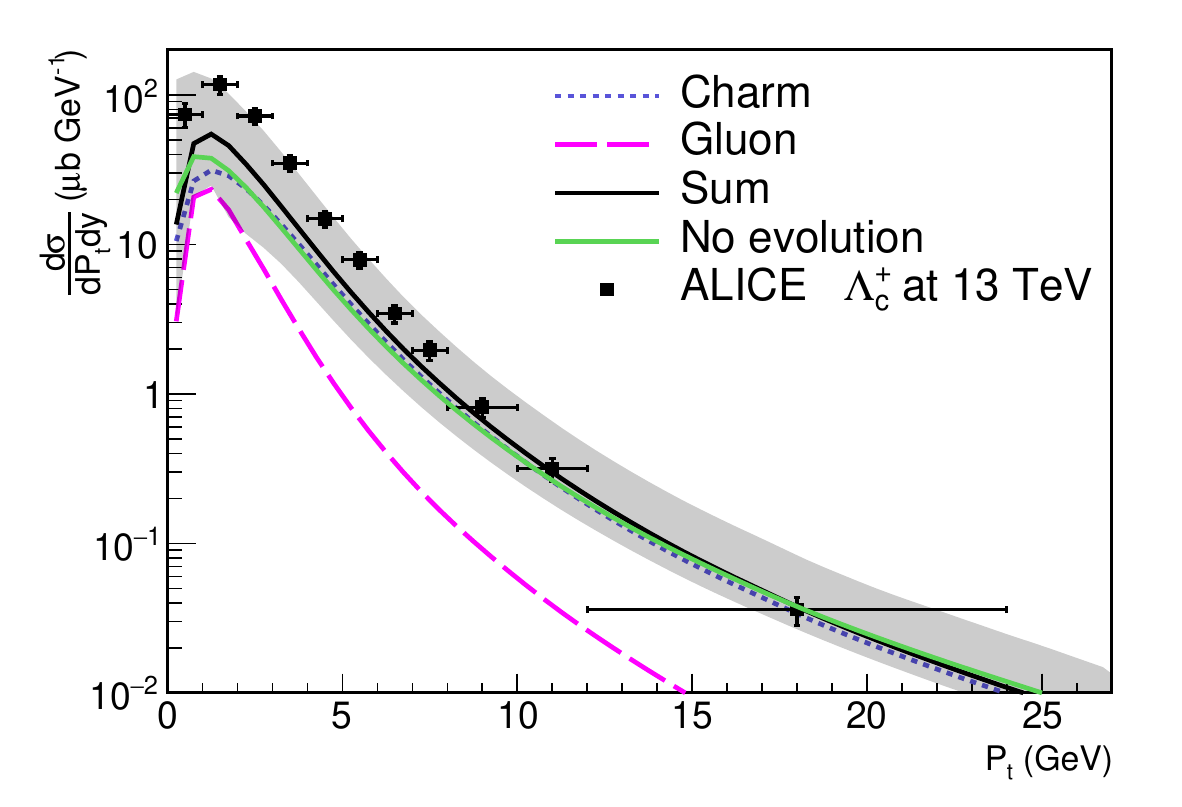}
\end{center}
\caption{Comparison between our calculations at 13 TeV in the rapidity range $|y|<0.5$ with ALICE data \cite{Acharya2022}. The green curve represents the complete contribution without evolution of the FF and $\epsilon_{\Lambda_c}=0.05$. \label{lcAlice2022}}
\end{figure}

More recently, the ALICE collaboration released new results, with measurements of $\Lambda_c^+$, $\Sigma_c^{0}$, $\Sigma_c^{++}$, $\Xi_c^{0}$, and $\Xi_c^{+}$ particles at 13 TeV \cite{Acharya2021,Acharya2022,Acharya2023}. We show our result for the $\Lambda_c^+$ baryon in Fig.~\ref{lcAlice2022}.
We observe a good  agreement for $P_t>5$ GeV and an underestimation of about a factor of 3 at small transverse momentum. The green curve represents the total contribution without evolution (we changed the FF parameter to $\epsilon_{\Lambda_c}=0.05$). Again, we observe that the gluon contribution helps to describe small-$P_t$ data. One may argue that the necessity for a new mechanism at small $P_t$ is unclear. Indeed, Fig.~\ref{lcAlice2022} does not look fundamentally different from $D$ meson results, with the usual underestimation at small transverse momentum, see, for instance, the top-left panel of Fig. 9 in \cite{Acharya2019} for the comparison between ALICE data and FONLL \cite{Cacciari1998,Cacciari2001} calculations. However, experimental ratios of $\Lambda_c/D$ \cite{Acharya2023} and $\Lambda_b/B$ \cite{Aaij2024a} indicate a clear enhancement of baryon production at small $P_t$. Several mechanisms have been proposed, but it goes beyond the scope of this work as they are, in general, not related to FFs. We note, however, that specific relations between the heavy-quark and hadron rapidities may result in enhanced production of baryons compared to mesons \cite{Maciula2018}. Note also that Refs. \cite{Acharya2022,Acharya2023} did not compare their measurements to pQCD calculations based on the collinear factorization. We can find a comparison with PYTHIA8 (Monash 2013), based on string fragmentation, that underestimates the ratio $\Lambda_c^+/D^0$ by a factor of 4 to 10 at small $P_t$, while our calculations would give approximately a factor of 2 to 4 (see also Ref.~\cite{Kniehl2020}).\\

\begin{figure}[h!]
\begin{center}
 \includegraphics[width=24pc]{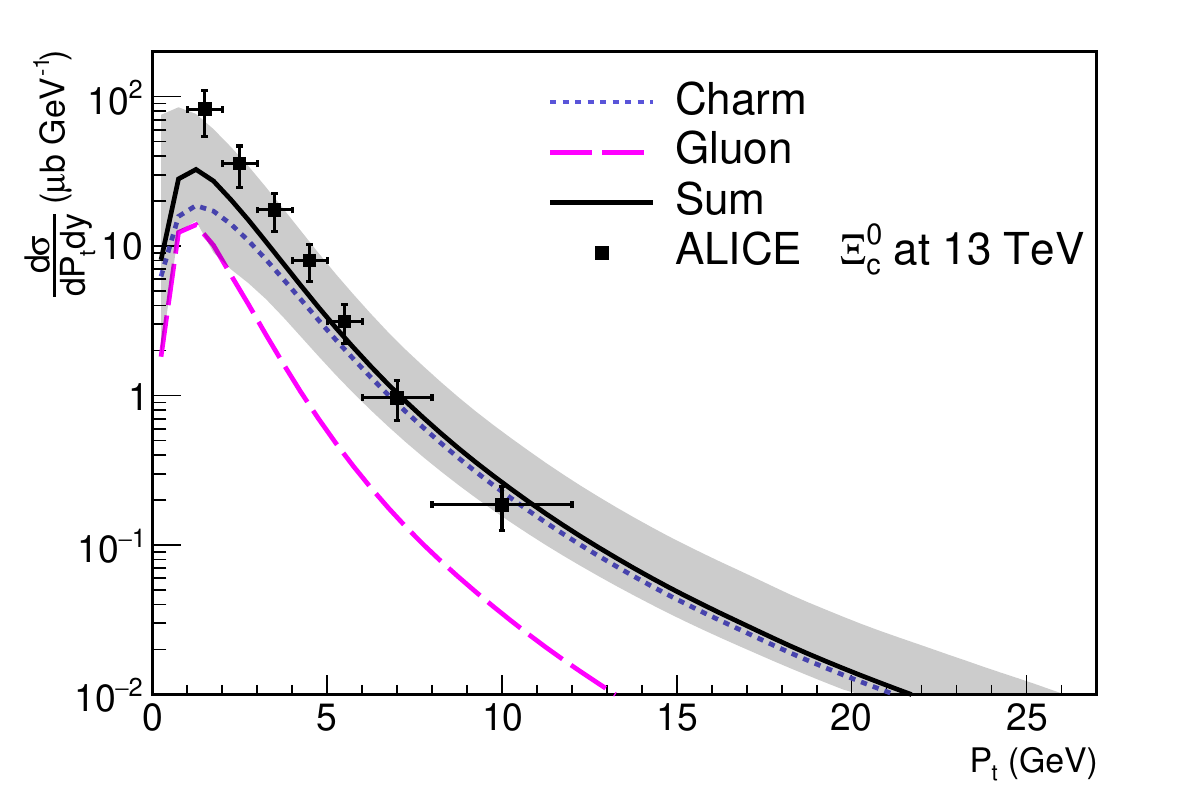}
\end{center}
\caption{Comparison between our calculations at 13 TeV in the rapidity range $|y|<0.5$ with ALICE data \cite{Acharya2021}.\label{xi0Alice2021}}
\end{figure}

\begin{figure}[h!]
\begin{center}
 \includegraphics[width=24pc]{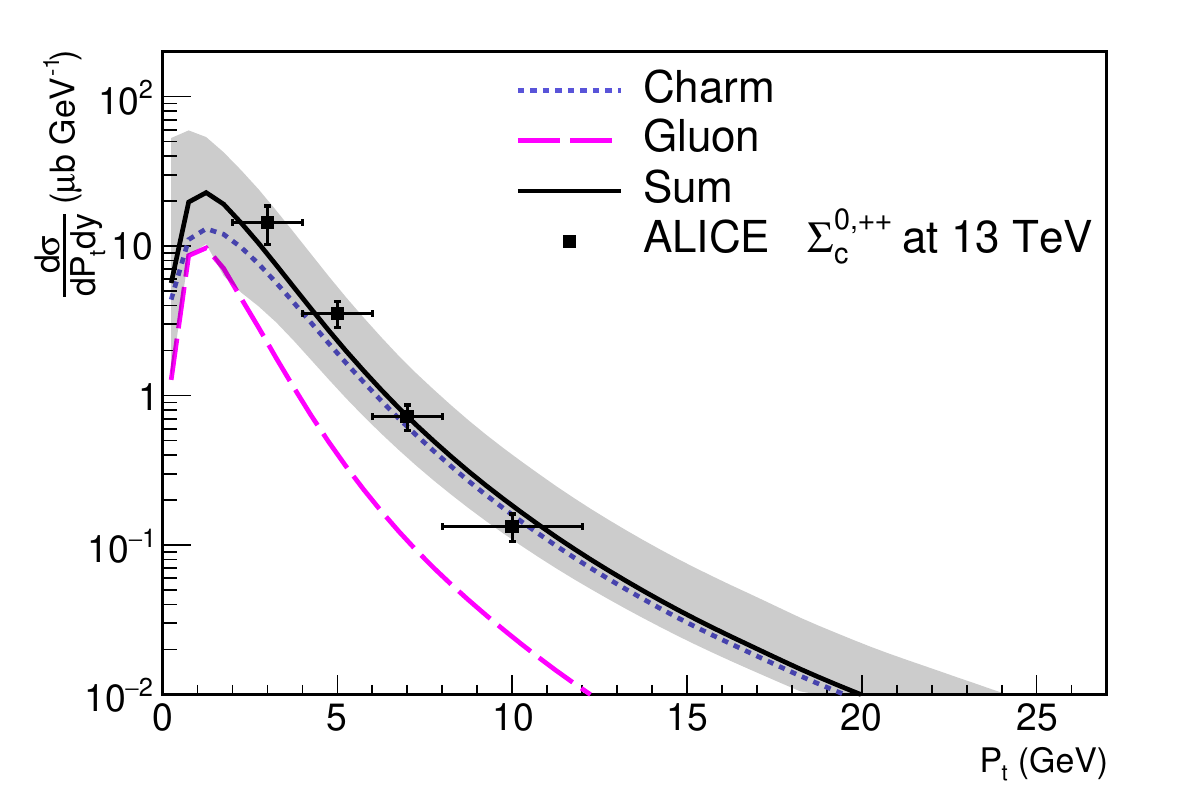}
\end{center}
\caption{Comparison between our calculations at 13 TeV in the rapidity range $|y|<0.5$ with ALICE data \cite{Acharya2022}.\label{Sig_cAlice2021}}
\end{figure}

We keep going with other charmed baryons. Ref.~\cite{Acharya2023} estimated several fragmentation fractions, giving the values $f_A(c\to \Lambda_c^{+})=0.168$, $f_A(c\to \Xi_c^{0})=0.099$, $f_A(c\to \Xi_c^{+})=0.096$ and  $f_A(c\to \Sigma_c^{0,+,++})=0.072$, where the subscript $A$ denotes the values determined by ALICE. However, the experimental definition of fragmentation fraction is not necessarily the same as the theoretical one since the former, based on $P_t$-integrated cross sections, includes any processes (except for some decays) leading to the desired particle, while the latter includes only the FF contribution and depends, for instance, on the order at which the partonic cross section is computed. Consequently, we fix our $\Xi_c^0$ fragmentation fraction to 
\begin{equation}
    f(c\to \Xi_c^{0})=f(c\to \Lambda_c^{+})\frac{f_A(c\to \Xi_c^{0})}{f_A(c\to \Lambda_c^{+})},
\end{equation}
with $f(c\to \Lambda_c^{+})=0.1$ used in this paper. This avoids introducing a new parameter.
The result obtained with the same setup as for $\Lambda_c^+$ baryons is presented in Fig.~\ref{xi0Alice2021}.
With the same procedure and
\begin{equation}
    f(c\to \Sigma_c^{0,++})=f(c\to \Lambda_c^{+})\frac{f_A(c\to \Sigma_c^{0,++})}{f_A(c\to \Lambda_c^{+})},
\end{equation}
we obtain the result for the $\Sigma_c^{0,++}$ baryons displayed in Fig.~\ref{Sig_cAlice2021}.\\

In conclusion, standard calculations using the Peterson FF describe charmed-baryon data, except at small transverse momentum. We did not show any results for bottomed baryons since the conclusion would be the same. We observed that the non-zero gluon-to-baryon contribution, due to the FF evolution, improves the agreement with small-$P_t$ data. The underestimation observed in this region does not question the universality of FFs but reveals that a new mechanism may be at work. We may, for instance, imagine describing the data using collinear factorization plus a coalescence model (similar to the one shown in Fig.~8 of Ref.~\cite{Acharya2023}), with the same energy-independent FFs from $e^+e^-$ (i.e., universal FFs) and the coalescence model filling the gap at small $P_t$. In the literature, the terminology of hadronization universality is sometimes used interchangeably with the FFs universality. However, it seems that the former does not have a precise mathematical definition and its breakdown seems to designate an unexpected hadron production in, e.g., pp collisions compared to $e^+e^-$. As exemplified above, it does not imply a breakdown of the universality of the fragmentation functions.

\section{The $B_c$ case \label{SecBc}}
%To close this paper, we present our result for the fragmentation contribution to the $B_c$ meson. (I just didn't know where to put this)

Among the $B$ mesons, the $B_c$ stands out as a unique case. It is the only meson made of two different heavy flavors, which makes its production and detection significantly more challenging than other b-flavored mesons and leads to significant uncertainties in several measurements of its properties. Consequently, instead of its differential production cross section, Ref.~\cite{Aaij2019} provides the transverse momentum dependence of its fragmentation ratio
\begin{equation} \label{FragmentationRatio}
    \frac{f_c}{f_u+f_d}
\end{equation}
with $f_c$, $f_u$, and $f_d$ the production fractions of the $B_c$, $B^+$, and $B^0$ mesons, respectively.\\

%Given the lack of precision in these measurements, fitting a fragmentation function for $B_c$ mesons will not yield predictions more accurate than the measurements they are compared to. Therefore, in this section, we adopt a theoretical approach to avoid these sources of error and attempt to constrain them further.\\

Our theoretical calculations include only the leading-power fragmentation contribution, with the factorization formula given by Eqs.~(\ref{ktfac}) and (\ref{facfrag}). We do not include higher-power contributions, with the short-distance production of the heavy-quark pair followed by its hadronization into quarkonia. The fragmentation functions of a parton into the $B_c$ meson can be computed within the NRQCD effective theory \cite{Caswell1986,Bodwin1995,Bodwin1997}. 
%Within the framework of NRQCD, non-relativistic bound states of heavy quarks, such as the $B_c$ meson, can be analyzed in a perturbative manner. Following a complementary factorization procedure which separates the production  of the heavy flavor from their bounding \cite{}, the fragmentation functions $D_{g\to B_c}$, $D_{c\to B_c}$, and $D_{b\to B_c}$ can be computed at some initial scale $\mu_{0}$. 
At leading order, the expressions for the color-singlet FFs can be obtained from \cite{Zheng2019} \footnote{Although the same authors published their results for the NLO FFs, they do not provide complete analytical expressions where we can set our own choice for the value of the heavy quark masses. Therefore, we use only the leading order expressions.}
\begin{align}
    D^{LO}_{b\rightarrow B_c}(z)=&\frac{2 \alpha^2_sz(1-z)^2 |R_s(0)|^2}{81\pi r_c^2(1-r_bz)^6M^3}[6-18(1-2r_c)z+(21-74r_c+68r_c^2)z^2 \nonumber \\
    &-2r_b(6-19r_c+18r_c^2)z^3+3r^2_b(1-2r_c+2r_c^2)z^4] \label{btoBc} \\ 
    D^{LO}_{b\rightarrow B_c^*}(z)=&\frac{2 \alpha^2_sz(1-z)^2 |R_s(0)|^2}{27\pi r_c^2(1-r_bz)^6M^3}[2-2(3-2r_c)z+3(3-2r_c+4r_c^2)z^2 \nonumber \\
    &-2r_b(4-r_c+2r_c^2)z^3+r^2_b(3-2r_c+2r_c^2)z^4] \label{btoBcp}
\end{align}
where $r_c = m_c/M$, $r_b=m_b/M$, and $M=m_b+m_c$, and $R_s(0)$ is the value of the $B_c$ wave function at the origin.  We replace the value of $|R_s(0)|^2$ given in \cite{Zheng2019} with $|R_s(0)|^2=6.21$ GeV$^3$ reported in \cite{Liao2014}. Indeed, Ref.~\cite{Baranov2018} pointed out that the value used in \cite{Zheng2019} yields predictions for $B_c$ production that are approximately a factor of four below the observations. The charm to $B_c$ FFs are obtained by interchanging $r_c$ and $r_b$ in Eqs.~(\ref{btoBc}) and (\ref{btoBcp}), while the gluon FF is the result of DGLAP evolution. \\

\begin{figure}[h!]
\begin{center}
 \includegraphics[width=21pc]{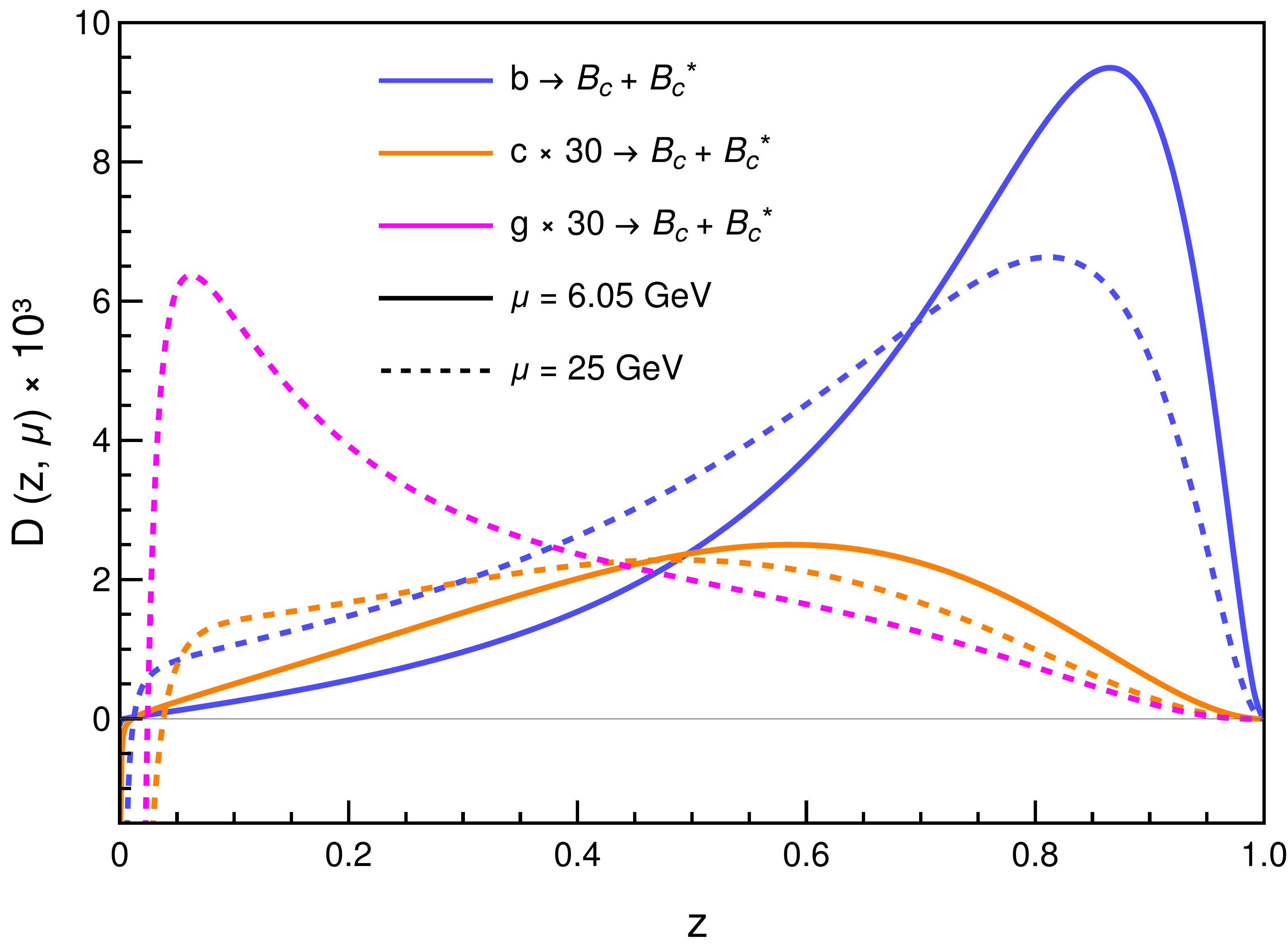}
\end{center}
\caption{LO FFs \cite{Zheng2019} for $B_c+B_c^*$ mesons. The gluon and charm FFs have been scaled by a factor of 30 to facilitate comparison with the bottom FF.}
\label{BcFFs}
\end{figure}

One of the main differences between the $B_c$ and $B$ hadrons lies in the fact that the mass of the former $B_c$ is significantly larger than the masses of individual b and c quarks. Moreover, the charm quark now has a nonzero FF at the initial scale. Following the logic of our earlier calculations, we choose
\begin{equation}
    \mu_{0}=m_{B_c}\simeq m_c+m_b.
\end{equation} 
We fix the fragmentation scale to
\begin{equation}
  \mu_F=\sqrt{P_t^2+m_{B_c}^2},
\end{equation}
while the factorization scale for charm and bottom production is left untouched. In Fig.~\ref{BcFFs}, we illustrate the behaviour of the gluon, charm, and bottom FFs at $\mu=\mu_0=6.05$ GeV (full lines) and $\mu=25$ GeV (dashed lines).
 Due to mass effects, the charm FF is smaller compared to the bottom quark. However, since charm quarks are produced more frequently than bottom quarks, this contribution cannot, a priori, be neglected. Note that Fig.~\ref{BcFFs} displays the results for $B_c+B_c^*$ because the $B_c^*$ meson decays essentially in $B_c$. Given that the mass difference between these two states is around 50 MeV, the momentum shift of a $B_c$ originating from a $B_c^*$ is negligible in the kinematic regime of LHCb. Thus, the total $B_c$ production cross section is the sum of the direct (without decays) production of $B_c$ and $B_c^*$.\\

\begin{figure}[h!]
\begin{center}
 \includegraphics[width=21pc]{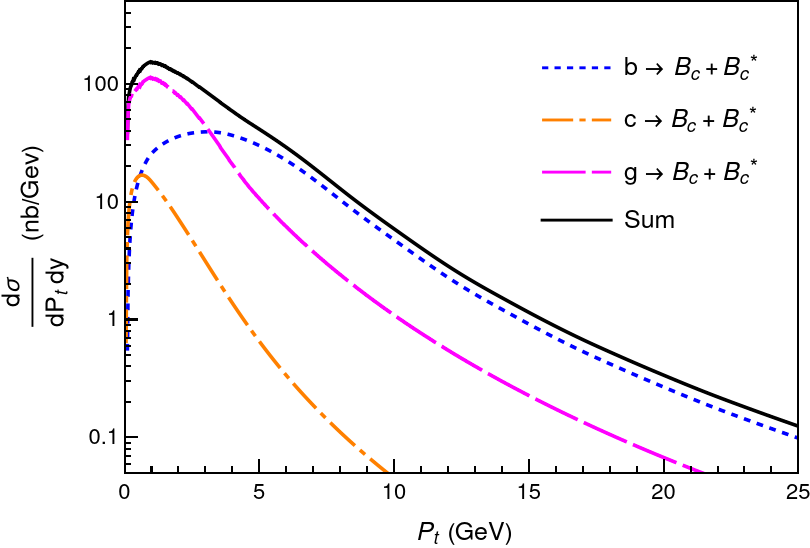}
\end{center}
\caption{Theoretical results for $B_c$ mesons obtained using the standard scheme, along with LO NRQCD predictions for their fragmentation functions.}
\label{BcStandar}
\end{figure}

%Before delving into our calculations, it is important to distinguish between what we can compute theoretically and what is observed in experiments. In experiments, only weakly decaying mesons are detected, meaning any hard or electromagnetic processes have already occurred. By contrast, using the NRQCD approach, our calculations give production rates of mesons at the moment of their creation. Since $B_c$ mesons belong to a family of flavored mesons, they are protected from annihilation, and their excited states, up to the second radial excitation \cite{Li2019}, have masses below the $BD$ production threshold. As a result, states such as $B_c^*$ ($^3S_1$) most of the time decay into the ground state $B_c$ ($^1S_0$) by electromagnetic ($B_c + \gamma$) or pionic radiation, significantly contributing to the number of weakly decaying $B_c$ mesons. Given that the mass difference between these two states is around 50 MeV \cite{Li2019}, the momentum shift of a $B_c$ originating from a $B_c^*$ is negligible in the kinematic regime of LHCb. Thus, the total $B_c$ production cross section can be considered as the sum of the direct (without decays) production of $B_c$ and $B_c^*$.\\
In Fig.~\ref{BcStandar}, we show our results for the $B_c$ production cross section using our default scheme.
 As could be expected, the gluon contribution is smaller than in the case of lighter B mesons. The reason for this is the higher initial scale, which slows down the evolution of the fragmentation functions, resulting in a smaller gluon FF. Additionally, we observe a negligible contribution from the charm fragmentation.\\

\begin{figure}[h!]
\begin{center}
 \includegraphics[width=21pc]{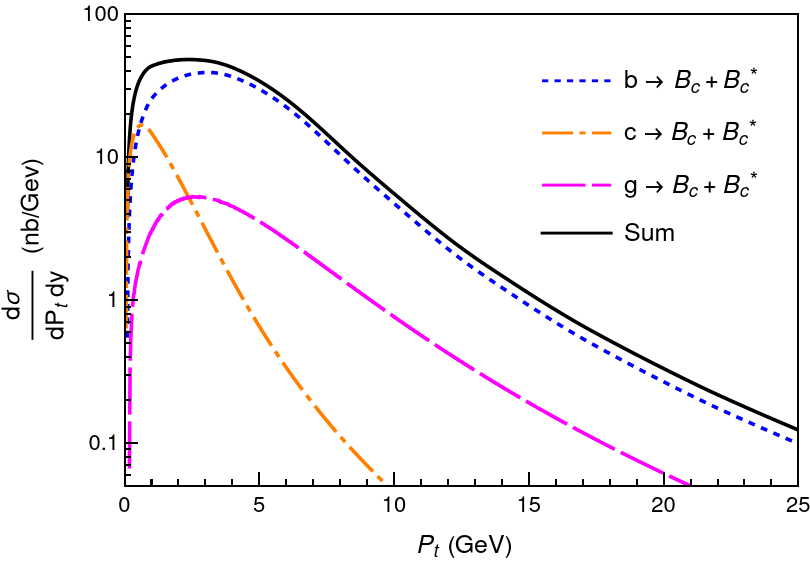}
\end{center}
\caption{Theoretical results for $B_c$ mesons obtained using the SACOT-$m_\T$ scheme, along with LO NRQCD predictions for their fragmentation functions.}
\label{BcSACOT}
\end{figure}

In order to apply the SACOT-$m_t$ scheme to double heavy mesons, we have to make an additional modification. As explained in section 3, this scheme is conceived to regularize the behavior of massless partons at small $p_t$ and  account for the virtuality of  gluons that split into heavy quarks present in the fragmentation functions. Since we are dealing with the specific scenario of a meson made of two different quarks, we need to consider the mass of both quarks. Therefore, instead of evaluating the gluon cross section at $\sqrt{p_t^2+M^2}$ with $M$ equal to $m_c$ or $m_b$,  we choose $M=m_{B_c}\approx m_c+m_b$. In Fig.~\ref{BcSACOT}, we present our results for the $B_c$ production cross section, now using the SACOT-$m_t$ scheme. 
The gluon contribution is now nearly negligible, a consequence of the shift in transverse momentum. That is a clear difference with the case of lighter B mesons.\\

\begin{figure}[h!]
\begin{center}
 \includegraphics[width=19pc]{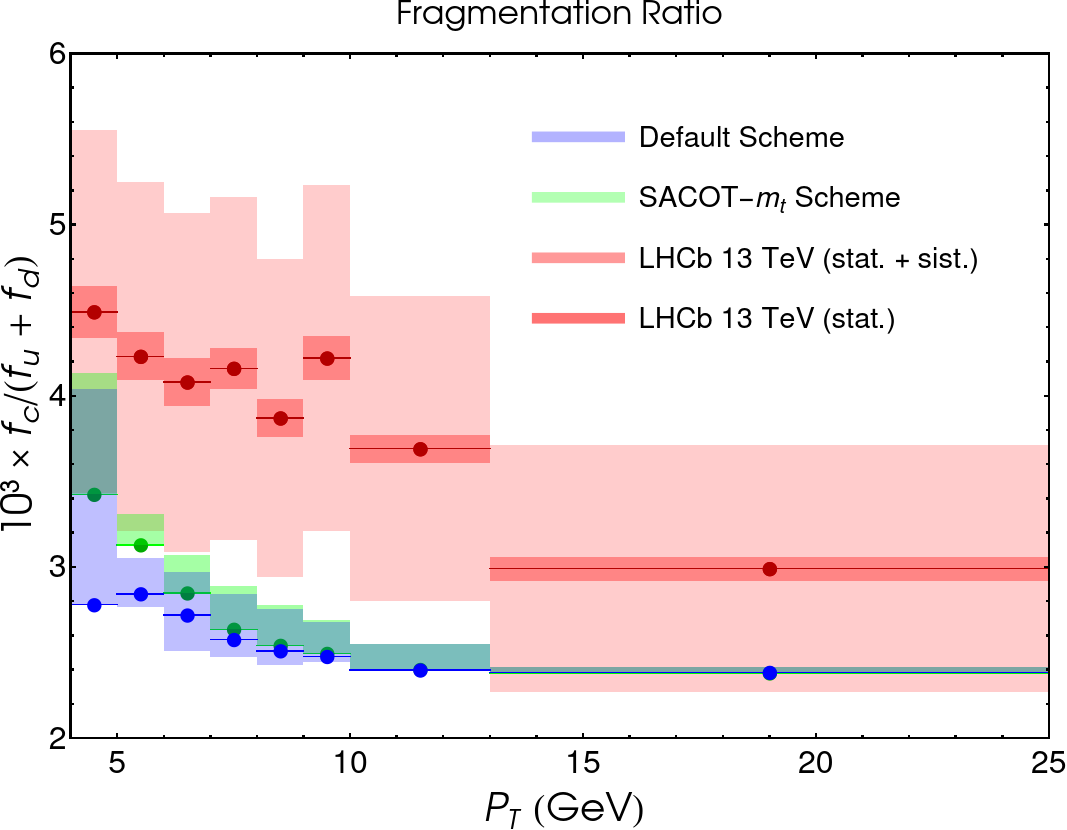}
\end{center}
\caption{Theoretical result for the fragmentation ratio $f_c/(f_u+f_d)$  compared to LHCb data \cite{Aaij2019}.}
\label{ratio}
\end{figure}

Finally, in Fig.~\ref{ratio}, we compare our results for the ratio \eqref{FragmentationRatio} with the LHCb measurement at $\sqrt{s}=$ 13 GeV and $2.5 < \eta < 4.5$ \cite{Aaij2019}.\footnote{We performed our calculations for $2<y<2.5$, which is not exactly the values used by the LHCb experiment. However, we expect the result to be the same because the LHCb collaboration saw no dependence on pseudo-rapidity.}
 Our calculations agree with data in the last bin but fail at smaller $P_t$. The steeper slope related to the SACOT-$m_t$ calculation can be explained by our earlier observation that, at small $P_t$, the gluon contribution to the $B_c$ production is smaller compared to light $B$ mesons. Since this scheme reduces the gluon contribution for that kinematic range, light $B$ mesons are more affected.\\

The underestimation of experimental data can, at least partially, be explained by missing contributions. Our calculations with the fragmentation formalism do not include the color-octet (CO) contribution, which is the true leading order in $\alpha_s$. However, the CO long-distance-matrix elements are generally small and cannot be extracted due to the lack of data. As already mentioned, we did not include either higher-power contributions where the heavy-quark pair is directly produced in the partonic cross section. In the case of, for instance, the $J/\psi$ quarkonium, these contributions dominate at small transverse momentum. We can consequently expect that the next-to-leading power terms will significantly improve the agreement with data at small $P_t$. The leading and next-to-leading powers have been discussed together in \cite{Kang2014} (and references therein), which explains, in particular, how to compute short distance coefficients to avoid double counting.\\

We note some similarities between the experimental ratio of Fig.~\ref{ratio} and the $\Lambda_b$ to $B$ meson ratio \cite{Aaij2024a}. In particular, the value of the ratios at $P_t=5$ GeV is about 1.5 times larger than at 20 GeV (while perturbative estimations are generally flat or increase slightly with $P_t$). Of course, it may be a coincidence, and these behaviors don't necessarily have the same origin. But since we expect an improvement in our calculations after including higher-power corrections to the $B_c$ cross section, we could wonder if a similar mechanism plays a role for the $\Lambda_b^0$ meson. Specifically, processes where two and three components of the $\Lambda_b^0$ baryon are produced at short distances are not considered. In that case, the following fragmentation functions would be required: $D_{[ub]\to \Lambda_b^0}$, $D_{[db]\to \Lambda_b^0}$, and $D_{[udb]\to \Lambda_b^0}$. NRQCD would not be applicable, but a similar formalism has been discussed by Braaten and collaborators for the fragmentation $D_{[s\bar{b}]\to B_s}$ \cite{Braaten2002}. This contribution was found to be significant at small $P_t$ and large rapidities.

\section{Conclusion \label{secdis}}
With this work, we completed the implementation of the HEF at order $\mathcal{O}(\alpha_s^2)$ within a full variable-flavor-number scheme and demonstrated the capability of this formalism to describe a large range of heavy-flavor observables. The DGLAP evolution of the FFs, managed by QCDNUM, renders the gluon-to-heavy-hadron contribution non-zero, improving the description of data at small $P_t$. It was also necessary to redetermine the non-perturbative parameters of the Peterson FF, which are systematically smaller compared to the case without evolution. Several FFs have been compared in Fig.~\ref{frag}, but we did not observe significant differences and worked with the Peterson FFs in the rest of the manuscript.\\

While the three schemes considered in this study work reasonably well for $D$ mesons, we saw that $B$ mesons seem to prefer the SACOT-$m_T$ and BKKSS schemes, while our default scheme is ruled out. We have a preference for SACOT-$m_T$ because in the BKKSS treatment, $\mu_f=P_t/2$ at large $P_t$. Then, this choice modifies the default scheme even in a region where mass effects are absent. However, the presence of large scale uncertainties at low $P_t$ and the limited transverse-momentum range of the data do not allow us to claim that one is better. With a scheme fixed once and for all, our HEF calculations accomplish the non-trivial task of describing quite accurately the whole set of heavy-hadron data selected for this study.\\

An exception to this statement is the production of heavy baryons at small $P_t$, which is usually underestimated by perturbative calculations, including ours. We saw, in particular from Fig.~\ref{lcAlice2022}, that the gluon contribution helps to reduce the gap between theory and experiment. This mismatch does not imply a violation of FFs universality but indicates that other mechanisms may be at work in this kinematical region. Indeed, at larger $P_t$, our calculations are in perfect agreement with data. Several non-perturbative models filling the gap at low $P_t$ are already on the market, and we mentioned at the end of Sec.~\ref{SecBc} that several perturbative contributions have not been considered yet.\\

In the last section, we made predictions for the transverse-momentum dependence of the $B_c$ meson cross section. We studied the contributions $D_{i\to B_c}$, where $i=g,c,b$, and the color-singlet FF is determined by applying the usual Feynman rules and NRQCD. In Fig.~{\ref{ratio}}, our calculation for the ratio $B_c/(B^0+B^+)$ underestimates the experimental result, particularly at small $P_t$. It didn't come as a surprise since higher-power contributions have not been included. Still, we note that the SACOT-$m_T$ scheme reproduces the shape better.

\section*{Acknowledgments}
We are grateful to Ted Rogers and Hannu Paukkunen for the discussions on factorization and fragmentation scales.
BG acknowledges support from ANID PIA/APOYO AFB230003. 

% BIBLIOGRAPHY
\bibliographystyle{BibFiles/t1}
\bibliography{BibFiles/GeneralBib}
\end{document}

%% file: Packages/myTikz.tex
\usepackage{tikz}

\usetikzlibrary{decorations.markings}
\usetikzlibrary{decorations.pathmorphing}

\tikzset{
    vector/.style={decorate, decoration={snake}, draw},
    fermion/.style={draw=black, postaction={decorate},
        decoration={markings,mark=at position .55 with {\arrow[scale=1.,>=stealth]{>}}}},
    fermionbar/.style={draw=black, postaction={decorate},
        decoration={markings,mark=at position .58 with {\arrow[scale=1.,>=stealth]{<}}}},
    fermionline/.style={draw=black},
    gluon/.style={decorate, draw=black,
        decoration={coil,amplitude=4pt, segment length=5pt}},
    scalar/.style={dashed, draw=black, postaction={decorate},
        decoration={markings,mark=at position .55 with {\arrow[scale=1.,>=stealth]{>}}}},
    scalarbar/.style={dashed, draw=black, postaction={decorate},
        decoration={markings,mark=at position .58 with {\arrow[scale=1.,>=stealth]{<}}}},
    scalarline/.style={dashed,draw=black},
    fmassin/.style={draw=black, postaction={decorate},
    decoration={markings, mark=at position .75 with {\arrow[scale=1.,>=stealth]{>}}, mark=at position .30 with {\arrow[scale=1.,>=stealth]{<}}}},
}

%% file: Packages/myHypref.tex
\usepackage[colorlinks=true,linkcolor=redLinks,citecolor=greenLinks,urlcolor=redLinks, pdfborder={0 0 1}]{hyperref}
\definecolor{greenLinks}{rgb}{0, 0.6, 0} 
\definecolor{blueLinks}{rgb}{0, 0, 0.6}
\definecolor{redLinks}{rgb}{0.6, 0, 0}
\definecolor{blackLinks}{rgb}{0.77,0.12,0.23} %not black
\definecolor{eprintLinks}{rgb}{0.4, 0.4, 0.4}
\definecolor{journalLinks}{rgb}{0.6, 0, 0}

%% file: main.bbl
\begin{thebibliography}{10}
\providecommand{\url}[1]{\texttt{#1}}
\providecommand{\urlprefix}{URL }
\providecommand{\eprint}[2][]{\url{#2}}

\bibitem{Guiot2021}
B.~Guiot and A.~van Hameren, \emph{${D}$ and ${B}$-meson production using
  $k_t$-factorization calculations in a variable-flavor-number scheme},
  \MYhref[journalLinks]{http://dx.doi.org/10.1103/physrevd.104.094038}{Phys.
  Rev. D
  }\MYhref[journalLinks]{http://dx.doi.org/10.1103/physrevd.104.094038}{\textbf{104}
  (2021) 9 094038}.

\bibitem{Catani1990}
S.~Catani, M.~Ciafaloni and F.~Hautmann, \emph{Gluon contributions to small $x$
  heavy flavour production},
  \MYhref[journalLinks]{http://dx.doi.org/10.1016/0370-2693(90)91601-7}{Phys.
  Lett. B
  }\MYhref[journalLinks]{http://dx.doi.org/10.1016/0370-2693(90)91601-7}{\textbf{242}
  (1990) 97--102}.

\bibitem{Catani1991}
S.~Catani, M.~Ciafaloni and F.~Hautmann, \emph{High energy factorization and
  small-x heavy flavour production},
  \MYhref[journalLinks]{http://dx.doi.org/10.1016/0550-3213(91)90055-3}{Nucl.
  Phys. B
  }\MYhref[journalLinks]{http://dx.doi.org/10.1016/0550-3213(91)90055-3}{\textbf{366}
  (1991) 1 135--188}.

\bibitem{Collins1991}
J.~Collins and R.~Ellis, \emph{Heavy-quark production in very high energy
  hadron collisions},
  \MYhref[journalLinks]{http://dx.doi.org/10.1016/0550-3213(91)90288-9}{Nucl.
  Phys. B
  }\MYhref[journalLinks]{http://dx.doi.org/10.1016/0550-3213(91)90288-9}{\textbf{360}
  (1991) 3--30}.

\bibitem{Levin91}
E.~M. Levin, M.~G. Ryskin, Y.~M. Shabelski and A.~G. Shuvaev, \emph{Heavy quark
  production in semihard nucleon interactions}, Soviet Journal of Nuclear
  Physics \textbf{53} (1991) 657.

\bibitem{Fadin1975}
V.~Fadin, E.~Kuraev and L.~Lipatov, \emph{On the {P}omeranchuk singularity in
  asymptotically free theories}, Phys. Lett. B \textbf{60} (1975) 50--52.

\bibitem{Kuraev1976}
E.~Kuraev, L.~Lipatov and V.~Fadin, \emph{Multi-{R}eggeon processes in the
  {Y}ang-{M}ills theory}, Sov. Phys. JETP \textbf{44} (1976) 443--450.

\bibitem{Kuraev1977}
E.~Kuraev, L.~Lipatov and V.~Fadin, \emph{The {P}omeranchuk singularity in
  nonabelian gauge theories}, Sov. Phys. JETP \textbf{45} (1977) 199--204.

\bibitem{Balitsky1978}
I.~Balitsky and L.~Lipatov, \emph{The {P}omeranchuk singularity in quantum
  chromodynamics}, Sov. J. Nucl. Phys. \textbf{28} (1978) 822--829.

\bibitem{Iancu2004}
E.~Iancu and R.~Venugopalan, \emph{T{HE COLOR GLASS CONDENSATE AND HIGH ENERGY
  SCATTERING IN QCD}},
  \MYhref[journalLinks]{http://dx.doi.org/10.1142/9789812795533_0005}{Quark-gluon
  plasma 4 }, pages 249--3363 (2003) .

\bibitem{Gelis2010}
F.~Gelis, E.~Iancu, J.~Jalilian-Marian and R.~Venugopalan, \emph{The {C}olor
  {G}lass {C}ondensate},
  \MYhref[journalLinks]{http://dx.doi.org/10.1146/annurev.nucl.010909.083629}{Ann.
  Rev. Nucl. Part. Sci.
  }\MYhref[journalLinks]{http://dx.doi.org/10.1146/annurev.nucl.010909.083629}{\textbf{60}
  (2010) 463}.

\bibitem{Kovchegov2022}
Y.~V. Kovchegov and E.~Levin, \emph{Quantum {C}hromodynamics at {H}igh
  {E}nergy}, Cambridge University Press (2022).

\bibitem{Kotko2015}
P.~Kotko et~al., \emph{Improved {TMD} factorization for forward dijet
  production in dilute-dense hadronic collisions},
  \MYhref[journalLinks]{http://dx.doi.org/10.1007/jhep09(2015)106}{JHEP
  }\MYhref[journalLinks]{http://dx.doi.org/10.1007/jhep09(2015)106}{\textbf{2015}
  (2015) 9}.

\bibitem{Boussarie2023}
R.~Boussarie et~al., \emph{T{MD} {H}andbook}  (2023),
  \MYhref[eprintLinks]{http://arxiv.org/abs/2304.03302}{{\ttfamily
  arXiv:2304.03302 [hep-ph]}}.

\bibitem{Guiot2019}
B.~Guiot, \emph{Heavy-quark production with $k_t$ factorization: {T}he
  importance of the sea-quark distribution},
  \MYhref[journalLinks]{http://dx.doi.org/10.1103/physrevd.99.074006}{Phys.
  Rev. D
  }\MYhref[journalLinks]{http://dx.doi.org/10.1103/physrevd.99.074006}{\textbf{99}
  (2019) 7 074006}.

\bibitem{Catani1994}
S.~Catani and F.~Hautmann, \emph{High-energy factorization and small-x deep
  inelastic scattering beyond leading order},
  \MYhref[journalLinks]{http://dx.doi.org/10.1016/0550-3213(94)90636-x}{Nucl.
  Phys. B
  }\MYhref[journalLinks]{http://dx.doi.org/10.1016/0550-3213(94)90636-x}{\textbf{427}
  (1994) 3 475--524}.

\bibitem{Peterson1983}
C.~Peterson, D.~Schlatter, I.~Schmitt and P.~M. Zerwas, \emph{Scaling
  violations in inclusive $e^+e^-$ annihilation spectra},
  \MYhref[journalLinks]{http://dx.doi.org/10.1103/physrevd.27.105}{Phys. Rev. D
  }\MYhref[journalLinks]{http://dx.doi.org/10.1103/physrevd.27.105}{\textbf{27}
  (1983) 105}.

\bibitem{Karpishkov2015}
A.~Karpishkov, M.~Nefedov, V.~Saleev and A.~Shipilova, \emph{Open charm
  production in the parton {R}eggeization approach: {T}evatron and the {LHC}},
  \MYhref[journalLinks]{http://dx.doi.org/10.1103/physrevd.91.054009}{Phys.
  Rev. D
  }\MYhref[journalLinks]{http://dx.doi.org/10.1103/physrevd.91.054009}{\textbf{91}
  (2015) 5 054009}.

\bibitem{Karpishkov2015a}
A.~V. Karpishkov, M.~A. Nefedov, V.~A. Saleev and A.~V. Shipilova,
  \emph{B-meson production in the {P}arton {R}eggeization {A}pproach at
  {T}evatron and the {LHC}},
  \MYhref[journalLinks]{http://dx.doi.org/10.1142/s0217751x15500232}{Int. J.
  Mod. Phys. A
  }\MYhref[journalLinks]{http://dx.doi.org/10.1142/s0217751x15500232}{\textbf{30}
  (2015) 1550023}.

\bibitem{Hameren2013}
A.~van Hameren, P.~Kotko and K.~Kutak, \emph{Helicity amplitudes for
  high-energy scattering},
  \MYhref[journalLinks]{http://dx.doi.org/10.1007/jhep01(2013)078}{JHEP
  }\MYhref[journalLinks]{http://dx.doi.org/10.1007/jhep01(2013)078}{\textbf{2013}
  (2013) 1}.

\bibitem{Hameren2018}
A.~van Hameren, \emph{{KaTie}: For parton-level event generation with
  $k_t$-dependent initial states},
  \MYhref[journalLinks]{http://dx.doi.org/10.1016/j.cpc.2017.11.005}{Computer
  Physics Communications
  }\MYhref[journalLinks]{http://dx.doi.org/10.1016/j.cpc.2017.11.005}{\textbf{224}
  (2018) 371--380}.

\bibitem{Kimber2001}
M.~A. Kimber, A.~D. Martin and M.~G. Ryskin, \emph{Unintegrated parton
  distributions},
  \MYhref[journalLinks]{http://dx.doi.org/10.1103/physrevd.63.114027}{Phys.
  Rev. D
  }\MYhref[journalLinks]{http://dx.doi.org/10.1103/physrevd.63.114027}{\textbf{63}
  (2001) 114027}.

\bibitem{Watt2003}
G.~Watt, A.~D. Martin and M.~G. Ryskin, \emph{Unintegrated parton distributions
  and inclusive jet productionat {HERA}},
  \MYhref[journalLinks]{http://dx.doi.org/10.1140/epjc/s2003-01320-4}{Eur.
  Phys. J. C
  }\MYhref[journalLinks]{http://dx.doi.org/10.1140/epjc/s2003-01320-4}{\textbf{31}
  (2003) 73--89}.

\bibitem{Guiot2023}
B.~Guiot, \emph{Normalization of unintegrated parton densities},
  \MYhref[journalLinks]{http://dx.doi.org/10.1103/physrevd.107.014015}{Phys.
  Rev. D
  }\MYhref[journalLinks]{http://dx.doi.org/10.1103/physrevd.107.014015}{\textbf{107}
  (2023) 1 014015}.

\bibitem{GribovLipatov:1972}
V.~N. Gribov and L.~N. Lipatov, \emph{Deep inelastic e p scattering in
  perturbation theory}, Sov. J. Nucl. Phys. \textbf{15} (1972) 438.

\bibitem{AltarelliParisi:1977}
G.~Altarelli and G.~Parisi, \emph{Asymptotic freedom in parton language},
  \MYhref[journalLinks]{http://dx.doi.org/10.1016/0550-3213(77)90384-4}{Nuclear
  Physics B
  }\MYhref[journalLinks]{http://dx.doi.org/10.1016/0550-3213(77)90384-4}{\textbf{126}
  (1977) 298--318}.

\bibitem{Dokshitzer:1977}
Y.~L. Dokshitzer, \emph{Calculation of the {S}tructure {F}unctions for {D}eep
  {I}nelastic {S}cattering and e+ e- {A}nnihilation by {P}erturbation {T}heory
  in {Q}uantum {C}hromodynamics.}, Sov. Phys. JETP \textbf{46} (1977) 641.

\bibitem{Botje2011}
M.~Botje, \emph{Q{CDNUM}: Fast {QCD} evolution and convolution},
  \MYhref[journalLinks]{http://dx.doi.org/10.1016/j.cpc.2010.10.020}{Comput.
  Phys. Commun.
  }\MYhref[journalLinks]{http://dx.doi.org/10.1016/j.cpc.2010.10.020}{\textbf{182}
  (2011) 490}.

\bibitem{Acharya2017}
S.~Acharya et~al., \emph{Measurement of {D}-meson production at mid-rapidity in
  pp collisions at $\sqrt{s}=7$ {T}e{V}},
  \MYhref[journalLinks]{http://dx.doi.org/10.1140/epjc/s10052-017-5090-4}{Eur.
  Phys. J. C
  }\MYhref[journalLinks]{http://dx.doi.org/10.1140/epjc/s10052-017-5090-4}{\textbf{77}
  (2017) 8}.

\bibitem{Collins1985}
P.~D.~B. Collins and T.~P. Spiller, \emph{The fragmentation of heavy quarks},
  \MYhref[journalLinks]{http://dx.doi.org/10.1088/0305-4616/11/12/006}{J. Phys.
  G
  }\MYhref[journalLinks]{http://dx.doi.org/10.1088/0305-4616/11/12/006}{\textbf{11}
  (1985) 1289}.

\bibitem{Kartvelishvili1978}
V.~Kartvelishvili, A.~Likhoded and V.~Petrov, \emph{On the fragmentation
  functions of heavy quarks into hadrons},
  \MYhref[journalLinks]{http://dx.doi.org/10.1016/0370-2693(78)90653-6}{Phy.
  Lett. B
  }\MYhref[journalLinks]{http://dx.doi.org/10.1016/0370-2693(78)90653-6}{\textbf{78}
  (1978) 615}.

\bibitem{Aaij2016}
R.~Aaij et~al., \emph{Measurements of prompt charm production cross-sections in
  pp collisions at $\sqrt{s} = 13$ {T}e{V}},
  \MYhref[journalLinks]{http://dx.doi.org/10.1007/jhep03(2016)159}{JHEP
  }\MYhref[journalLinks]{http://dx.doi.org/10.1007/jhep03(2016)159}{\textbf{2016}
  (2016) 3}.

\bibitem{Benzke2019}
M.~Benzke et~al., \emph{B-meson production in the general-mass
  variable-flavour-number scheme and {LHC} data},
  \MYhref[journalLinks]{http://dx.doi.org/10.1140/epjc/s10052-019-7326-y}{Eur.
  Phys. J. C
  }\MYhref[journalLinks]{http://dx.doi.org/10.1140/epjc/s10052-019-7326-y}{\textbf{79}
  (2019) 10}.

\bibitem{Helenius2018}
I.~Helenius and H.~Paukkunen, \emph{Revisiting the {D}-meson hadroproduction in
  general-mass variable flavour number scheme},
  \MYhref[journalLinks]{http://dx.doi.org/10.1007/jhep05(2018)196}{JHEP
  }\MYhref[journalLinks]{http://dx.doi.org/10.1007/jhep05(2018)196}{\textbf{2018}
  (2018) 5}.

\bibitem{Helenius2023}
I.~Helenius and H.~Paukkunen, \emph{B-meson hadroproduction in the
  {SACOT}-$m_{\text{t}}$ scheme},
  \MYhref[journalLinks]{http://dx.doi.org/10.1007/jhep07(2023)054}{JHEP
  }\MYhref[journalLinks]{http://dx.doi.org/10.1007/jhep07(2023)054}{\textbf{2023}
  (2023) 7}.

\bibitem{Cacciari2012}
M.~Cacciari et~al., \emph{Theoretical predictions for charm and bottom
  production at the {LHC}},
  \MYhref[journalLinks]{http://dx.doi.org/10.1007/jhep10(2012)137}{JHEP
  }\MYhref[journalLinks]{http://dx.doi.org/10.1007/jhep10(2012)137}{\textbf{2012}
  (2012) 10}.

\bibitem{Aaij2017b}
R.~Aaij et~al., \emph{Measurement of the ${B}^\pm$ production cross-section in
  pp collisions at $ \sqrt{s}=7$ and 13 {T}e{V}},
  \MYhref[journalLinks]{http://dx.doi.org/10.1007/jhep12(2017)026}{JHEP
  }\MYhref[journalLinks]{http://dx.doi.org/10.1007/jhep12(2017)026}{\textbf{2017}
  (2017) 12}.

\bibitem{Guiot2024}
B.~Guiot and A.~van Hameren, \emph{Examination of kt-factorization in a
  {Y}ukawa theory},
  \MYhref[journalLinks]{http://dx.doi.org/10.1007/jhep04(2024)085}{JHEP
  }\MYhref[journalLinks]{http://dx.doi.org/10.1007/jhep04(2024)085}{\textbf{2024}
  (2024) 4}.

\bibitem{Kniehl2020}
B.~Kniehl, G.~Kramer, I.~Schienbein and H.~Spiesberger,
  \emph{{$\Lambda$}$_c^\pm$ production in $pp$ collisions with a new
  fragmentation function},
  \MYhref[journalLinks]{http://dx.doi.org/10.1103/physrevd.101.114021}{Phys.
  Rev. D
  }\MYhref[journalLinks]{http://dx.doi.org/10.1103/physrevd.101.114021}{\textbf{101}
  (2020) 11 114021}.

\bibitem{Aaij2013}
R.~Aaij et~al., \emph{Prompt charm production in pp collisions at $\sqrt{s}=7$
  {T}e{V}},
  \MYhref[journalLinks]{http://dx.doi.org/10.1016/j.nuclphysb.2013.02.010}{Nucl.
  Phys. B
  }\MYhref[journalLinks]{http://dx.doi.org/10.1016/j.nuclphysb.2013.02.010}{\textbf{871}
  (2013) 1}.

\bibitem{Sirunyan2020}
A.~Sirunyan et~al., \emph{Production of {$\Lambda$}$_c^+$ baryons in
  proton-proton and lead-lead collisions at $\sqrt{s_{\mathrm{NN}}}=5.02$
  {T}e{V}},
  \MYhref[journalLinks]{http://dx.doi.org/10.1016/j.physletb.2020.135328}{Phys.
  Lett. B
  }\MYhref[journalLinks]{http://dx.doi.org/10.1016/j.physletb.2020.135328}{\textbf{803}
  (2020) 135328}.

\bibitem{Acharya2018}
S.~Acharya et~al., \emph{{$\Lambda$}$_c^+$ production in pp collisions at
  $\sqrt{s}=7$ {T}e{V} and in p-{P}b collisions at
  $\sqrt{s_{\mathrm{NN}}}=5.02$ {T}e{V}},
  \MYhref[journalLinks]{http://dx.doi.org/10.1007/jhep04(2018)108}{JHEP
  }\MYhref[journalLinks]{http://dx.doi.org/10.1007/jhep04(2018)108}{\textbf{2018}
  (2018) 4}.

\bibitem{Maciula2018}
R.~Maciu{\l}a and A.~Szczurek, \emph{Production of {$\Lambda$}$_c$ baryons at
  the {LHC} within the $k_t$-factorization approach and independent parton
  fragmentation picture},
  \MYhref[journalLinks]{http://dx.doi.org/10.1103/physrevd.98.014016}{Phys.
  Rev. D
  }\MYhref[journalLinks]{http://dx.doi.org/10.1103/physrevd.98.014016}{\textbf{98}
  (2018) 014016}.

\bibitem{Cacciari2003}
M.~Cacciari and P.~Nason, \emph{Charm cross sections for the {T}evatron {R}un
  {II}},
  \MYhref[journalLinks]{http://dx.doi.org/10.1088/1126-6708/2003/09/006}{JHEP
  }\MYhref[journalLinks]{http://dx.doi.org/10.1088/1126-6708/2003/09/006}{\textbf{2003}
  (2003) 09}.

\bibitem{Tumasyan2024}
A.~Tumasyan et~al., \emph{Study of charm hadronization with prompt
  ${\Lambda}$$_c^+$ baryons in proton-proton and lead-lead collisions at
  $\sqrt{s_{\textrm{NN}}} = 5.02$ {T}e{V}},
  \MYhref[journalLinks]{http://dx.doi.org/10.1007/jhep01(2024)128}{JHEP
  }\MYhref[journalLinks]{http://dx.doi.org/10.1007/jhep01(2024)128}{\textbf{2024}
  (2024) 1}.

\bibitem{Christiansen2015}
J.~R. Christiansen and P.~Z. Skands, \emph{String formation beyond leading
  colour},
  \MYhref[journalLinks]{http://dx.doi.org/10.1007/jhep08(2015)003}{JHEP
  }\MYhref[journalLinks]{http://dx.doi.org/10.1007/jhep08(2015)003}{\textbf{2015}
  (2015) 8}.

\bibitem{Acharya2021}
S.~Acharya et~al., \emph{{M}easurement of the {C}ross {S}ections of
  {$\Xi$}$_c^0$ and {$\Xi$}$_c^+$ {B}aryons and of the {B}ranching-{F}raction
  {R}atio ${BR}(${$\Xi$}$_c^0 \to${$\Xi$}$^- e^+ \nu_e)/{BR}(${$\Xi$}$_c^0 \to$
  {$\Xi$}$^- \pi^+)$ in $pp$ {C}ollisions at $\sqrt{s}=13$ {T}e{V}},
  \MYhref[journalLinks]{http://dx.doi.org/10.1103/physrevlett.127.272001}{Phys.
  Rev. Lett.
  }\MYhref[journalLinks]{http://dx.doi.org/10.1103/physrevlett.127.272001}{\textbf{127}
  (2021) 27}.

\bibitem{Acharya2022}
S.~Acharya et~al., \emph{Measurement of prompt ${D}^0$, {$\Lambda$}$_c^+$, and
  {$\Sigma$}$_c^{0,++}$ (2455) production in proton-proton collisions at
  $\sqrt{s}=13$ {T}e{V}},
  \MYhref[journalLinks]{http://dx.doi.org/10.1103/physrevlett.128.012001}{Phys.
  Rev. Lett.
  }\MYhref[journalLinks]{http://dx.doi.org/10.1103/physrevlett.128.012001}{\textbf{128}
  (2022) 1, 012001}.

\bibitem{Acharya2023}
S.~Acharya et~al., \emph{Charm production and fragmentation fractions at
  midrapidity in pp collisions at $ \sqrt{s} = 13$ {T}e{V}},
  \MYhref[journalLinks]{http://dx.doi.org/10.1007/jhep12(2023)086}{JHEP
  }\MYhref[journalLinks]{http://dx.doi.org/10.1007/jhep12(2023)086}{\textbf{2023}
  (2023) 12}.

\bibitem{Acharya2019}
S.~Acharya et~al., \emph{Measurement of ${D}^0$ , ${D}^+$ , ${D}^{*+}$ and
  ${D}^+_s$ production in pp collisions at $\sqrt{s} = 5.02$ {T}e{V} with
  {ALICE}},
  \MYhref[journalLinks]{http://dx.doi.org/10.1140/epjc/s10052-019-6873-6}{Eur.
  Phys. J. C
  }\MYhref[journalLinks]{http://dx.doi.org/10.1140/epjc/s10052-019-6873-6}{\textbf{79}
  (2019) 5}.

\bibitem{Cacciari1998}
M.~Cacciari, M.~Greco and P.~Nason, \emph{The p$_{T}$ spectrum in heavy-flavour
  hadroproduction},
  \MYhref[journalLinks]{http://dx.doi.org/10.1088/1126-6708/1998/05/007}{JHEP
  }\MYhref[journalLinks]{http://dx.doi.org/10.1088/1126-6708/1998/05/007}{\textbf{1998}
  (1998) 007}.

\bibitem{Cacciari2001}
M.~Cacciari, S.~Frixione and P.~Nason, \emph{The p$_{T}$ spectrum in
  heavy-flavour photoproduction},
  \MYhref[journalLinks]{http://dx.doi.org/10.1088/1126-6708/2001/03/006}{JHEP
  }\MYhref[journalLinks]{http://dx.doi.org/10.1088/1126-6708/2001/03/006}{\textbf{2001}
  (2001) 006}.

\bibitem{Aaij2024a}
R.~Aaij et~al., \emph{Enhanced {P}roduction of {$\Lambda$}$_b^0$ {B}aryons in
  {H}igh-{M}ultiplicity $pp$ {C}ollisions at $\sqrt{s} = 13$ {T}e{V}},
  \MYhref[journalLinks]{http://dx.doi.org/10.1103/physrevlett.132.081901}{Phys.
  Rev. Lett.
  }\MYhref[journalLinks]{http://dx.doi.org/10.1103/physrevlett.132.081901}{\textbf{132}
  (2024) 081901}.

\bibitem{Aaij2019}
R.~Aaij et~al. (LHCb), \emph{{Measurement of the $B_c^-$ meson production
  fraction and asymmetry in 7 and 13 TeV $pp$ collisions}},
  \MYhref[journalLinks]{http://dx.doi.org/10.1103/PhysRevD.100.112006}{Phys.
  Rev. D
  }\MYhref[journalLinks]{http://dx.doi.org/10.1103/PhysRevD.100.112006}{\textbf{100}
  (2019) 11 112006},
  \MYhref[eprintLinks]{http://arxiv.org/abs/1910.13404}{{\ttfamily
  arXiv:1910.13404 [hep-ex]}}.

\bibitem{Caswell1986}
W.~Caswell and G.~Lepage, \emph{Effective lagrangians for bound state problems
  in {QED}, {QCD}, and other field theories},
  \MYhref[journalLinks]{http://dx.doi.org/10.1016/0370-2693(86)91297-9}{Phys.
  Lett. B
  }\MYhref[journalLinks]{http://dx.doi.org/10.1016/0370-2693(86)91297-9}{\textbf{167}
  (1986) 437--442}.

\bibitem{Bodwin1995}
G.~T. Bodwin, E.~Braaten and G.~P. Lepage, \emph{Rigorous {QCD} analysis of
  inclusive annihilation and production of heavy quarkonium},
  \MYhref[journalLinks]{http://dx.doi.org/10.1103/physrevd.51.1125}{Phys. Rev.
  D
  }\MYhref[journalLinks]{http://dx.doi.org/10.1103/physrevd.51.1125}{\textbf{51}
  (1995) 1125}.

\bibitem{Bodwin1997}
G.~T. Bodwin, E.~Braaten and G.~P. Lepage, \emph{Erratum: {R}igorous {QCD}
  analysis of inclusive annihilation and production of heavy quarkonium
  [{P}hys. {R}ev. d51, 1125 (1995)]},
  \MYhref[journalLinks]{http://dx.doi.org/10.1103/physrevd.55.5853}{Phys. Rev.
  D
  }\MYhref[journalLinks]{http://dx.doi.org/10.1103/physrevd.55.5853}{\textbf{55}
  (1997) 5853}.

\bibitem{Zheng2019}
X.-C. Zheng, C.-H. Chang, T.-F. Feng and X.-G. Wu, \emph{Q{CD} {NLO}
  fragmentation functions for $c$ or $b^-$ quark to ${B}_c$ or ${B}_c^*$ meson
  and their application},
  \MYhref[journalLinks]{http://dx.doi.org/10.1103/physrevd.100.034004}{Phys.
  Rev. D
  }\MYhref[journalLinks]{http://dx.doi.org/10.1103/physrevd.100.034004}{\textbf{100}
  (2019) 034004}.

\bibitem{Liao2014}
Q.-L. Liao and G.-Y. Xie, \emph{{Heavy quarkonium wave functions at the origin
  and excited heavy quarkonium production via top quark decays at the LHC}},
  \MYhref[journalLinks]{http://dx.doi.org/10.1103/PhysRevD.90.054007}{Phys.
  Rev. D
  }\MYhref[journalLinks]{http://dx.doi.org/10.1103/PhysRevD.90.054007}{\textbf{90}
  (2014) 5 054007},
  \MYhref[eprintLinks]{http://arxiv.org/abs/1408.5563}{{\ttfamily
  arXiv:1408.5563 [hep-ph]}}.

\bibitem{Baranov2018}
S.~P. Baranov and A.~V. Lipatov, \emph{{First estimates of the $B_c$ wave
  function from the data on the $B_c$ production cross section}},
  \MYhref[journalLinks]{http://dx.doi.org/10.1016/j.physletb.2018.09.007}{Phys.
  Lett. B
  }\MYhref[journalLinks]{http://dx.doi.org/10.1016/j.physletb.2018.09.007}{\textbf{785}
  (2018) 338--341},
  \MYhref[eprintLinks]{http://arxiv.org/abs/1805.05390}{{\ttfamily
  arXiv:1805.05390 [hep-ph]}}.

\bibitem{Kang2014}
Z.-B. Kang, Y.-Q. Ma, J.-W. Qiu and G.~Sterman, \emph{Heavy quarkonium
  production at collider energies: {F}actorization and evolution},
  \MYhref[journalLinks]{http://dx.doi.org/10.1103/physrevd.90.034006}{Phys.
  Rev. D
  }\MYhref[journalLinks]{http://dx.doi.org/10.1103/physrevd.90.034006}{\textbf{90}
  (2014) 034006}.

\bibitem{Braaten2002}
E.~Braaten, Y.~Jia and T.~Mehen, \emph{B production asymmetries in perturbative
  {QCD}},
  \MYhref[journalLinks]{http://dx.doi.org/10.1103/physrevd.66.034003}{Phys.
  Rev. D
  }\MYhref[journalLinks]{http://dx.doi.org/10.1103/physrevd.66.034003}{\textbf{66}
  (2002) 034003}.

\end{thebibliography}
